\documentclass[lettersize,journal]{IEEEtran}

\usepackage[caption=false,font=normalsize,labelfont=sf,textfont=sf]{subfig}
\usepackage{textcomp}
\usepackage{url}
\usepackage{verbatim}
\usepackage{enumitem}

\usepackage{times} 
\usepackage{graphicx}
\usepackage{cite}
\usepackage{citesort}
\usepackage{color}
\usepackage{psfrag}
\usepackage{caption}
\usepackage{amssymb}
\usepackage{stfloats}
\usepackage{epsfig}
\usepackage{pifont}
\usepackage{amsmath}

\usepackage{cases}
\usepackage{array}
\usepackage{multicol}
\usepackage{multirow}
\usepackage[T1]{fontenc}
\usepackage{comment}
\usepackage{makecell}
\usepackage{amsthm}
\usepackage{indentfirst}
\usepackage{setspace}
\usepackage{bm}
\usepackage{float}
\usepackage{cases}
\usepackage[ruled, lined, linesnumbered, commentsnumbered, longend]{algorithm2e}
\SetKwInOut{KwIn}{Input}
\SetKwInOut{KwOut}{Output}
\SetKwProg{Init}{[Initialization]}{}{end}
\SetKwProg{QT}{[Applying the ELA method]}{}{end}
\SetKwProg{DT}{[Outer Layer Iteration]}{}{end}
\SetKwProg{LT}{[Inner Layer Iteration]}{}{end}
\SetKwProg{QB}{[Iteratively implementing the optimization at each RF]}{}{end}
\graphicspath{{figures/}}
\theoremstyle{remark}
\newtheorem{remark}{Remark}
\makeatletter
\renewcommand{\maketag@@@}[1]{\hbox{\m@th\normalsize\normalfont#1}}
\makeatother

\renewenvironment{thebibliography}[1]{
	\begin{oldthebibliography}{#1}
		\setlength{\itemsep}{0em}
		\setlength{\parskip}{-0.1em}
	}
	{
	\end{oldthebibliography}
}

\makeatletter
\def\ifundefined{\@ifundefined}

\begin{document}
	
	\title{Sensing-Assisted Predictive Beamforming for UAV-Enabled Ocean Monitoring Networks}
	
	\author{Bohan Li,~\textit{Member, IEEE}, Guangfei Gao, Jinpeng Zhang, Min Ye, Qian Li, Huaming Yu, Jingjing Wang,~\textit{Senior Member, IEEE}, Pei Xiao,~\textit{Senior Member, IEEE}, 	Sheng Chen,~\IEEEmembership{Life Fellow, IEEE} %
		\thanks{This work was supported in part by the National Key Research and Development Program of China under Grant 2024YFC3109100, in part by the National Natural Science Foundation of China under Grant 62401530, in part by the Natural Science Foundation of Shandong Province under Grant 2026HWYQ-027 and ZR2024QF076, and in part by the National Defense Basic Scientific Research Program of China under Grant JCKY2024210C61424030101. ({\em Corresponding author: Min Ye.})} 
		\thanks{B. Li, G. Gao, M. Ye, and Q. Li are with the Faculty of Information Science and Engineering, the Engineering Research Center of Advanced Marine Physical Instruments and Equipment (Ministry of Education), and Qingdao Key Laboratory of Optics and Optoelectronics, Ocean University of China, Qingdao 266100, China (emails: bohan.li@ouc.edu.cn, ggf@stu.ouc.edu.cn, yemin@ouc.edu.cn, liqian@ouc.edu.cn).} %
	\thanks{H. Yu is with 
			the College of Oceanic and Atmospheric Sciences, Ocean University of China, Qingdao 266100, China (email:  hmyu@ouc.edu.cn)}
	\thanks{J. Zhang is with the National Key Laboratory of Electromagnetic Environment,
		China Research Institute of Radiowave Propagation, Qingdao 266107,
		China (email: zhangjp@crirp.ac.cn).}
	\thanks{J. Wang is with School of Cyber Science and Technology, Beihang
		University, Beijing 100191, China (email: drwangjj@buaa.edu.cn)}
	\thanks{P. Xiao is with 5GIC \& 6GIC, University of Surrey, Guildford GU2 7XH, U.K. (email: p.xiao@surrey.ac.uk)} %
    \thanks{S. Chen is with School of Electronics and Computer Science, University of Southampton, Southampton SO17 1BJ, U.K., and also with the Faculty of Information Science and Engineering, Ocean University of China, Qingdao 266100, China (email: sqc@ecs.soton.ac.uk).}
		\vspace*{-5mm}
	}
	
	\maketitle
	
	\begin{abstract}
		Surface buoys in ocean monitoring networks undergo wave-induced motion and current-induced drift, while  sea clutter further degrades sensing reliability; together, these effects challenge predictive beam alignment and uplink communication in UAV-enabled data collection. This paper investigates a sensing-assisted predictive beamforming framework for UAV--buoy maritime monitoring by explicitly accounting for wave-induced buoy dynamics and residual sea clutter. A frame-based UAV mission workflow is first established, where the UAV transmits integrated sensing and communication signals to acquire buoy echoes and to support subsequent uplink beam alignment. To characterize short-horizon buoy motion, a correlated-acceleration state-space model is developed by combining a Singer process for wave-driven excitation with a slowly varying current-drift term. Given the resulting nonlinear reflection, Doppler, and delay measurements, the posterior Fisher information matrix and the corresponding posterior Cram\'er--Rao bound (PCRB) are derived, and the predicted horizontal-position PCRB is adopted as the sensing metric. A per-frame worst-buoy design is then formulated to jointly optimize sensing power allocation and UAV position under uplink-rate, UAV-power, and mobility constraints. By exploiting a Schur-complement reformulation and a lagged successive convex approximation, the resulting subproblem is converted into a convex conic program with tractable complexity. Simulation results show that the proposed scheme maintains robust prediction and communication performance under denser buoy deployments and harsher sea conditions, and outperforms several baseline designs. In particular, the pronounced root mean square error (RMSE) degradation of the communication-only benchmark confirms that sensing-assisted state refinement is essential for accurate predictive beamforming in dynamic maritime environments. Compared with a full first-order Taylor expansion method, it achieves a more attractive performance--complexity tradeoff for online deployment.
	\end{abstract}

	
	\vspace*{-1mm}
	\section{Introduction}\label{S1}
	
    Ocean monitoring networks are fundamental to a wide range of maritime applications, including ecological surveillance, climate observation, disaster warning, and offshore operation management \cite{lin2020ocean}. In practice, a common and cost-effective approach is to deploy a large number of lightweight, battery-powered surface buoys that continuously sample local environmental parameters \cite{xia2020maritime}. However, beyond coastal coverage, delivering these sensing data to a remote control center is challenging due to the absence of stable infrastructure backhaul and the stringent energy budget of the buoys \cite{albaladejo2010wireless}. 

	Against this backdrop, unmanned aerial vehicles (UAVs) have emerged as a flexible data-collection platform \cite{Zeng2019UAVSurvey}. In particular, by flying over a designated sea area, a UAV can establish line-of-sight (LoS) links to dispersed buoys, retrieve buffered measurements, and further relay the data to shore or to a satellite gateway, thereby enabling on-demand maritime backhaul with fast deployment and wide-area coverage \cite{li2020maritime}.

	Despite the above advantages, UAV--buoy communications face a unique maritime impairment, namely, surface buoys are continuously driven by sea waves and currents, which induce non-negligible random motion and causes rapid variations in the relative range and angles \cite{fossen2021handbook}. This mobility is particularly detrimental when the UAV employs a large antenna array and highly directional beams to compensate for path loss and to support high-rate uplink transmissions. Even a small pointing mismatch can lead to a severe beamforming gain loss, resulting in unstable data rates and frequent link outages. Moreover, wave-induced sea-surface reflections may introduce sea clutter into the UAV's sensing echoes, further degrading the accuracy of buoy-state acquisition \cite{ward2006sea,Huang2021SeaClutter}. Therefore, to ensure robust data collection, the UAV must be able to sense, track, and, more importantly, predict short-term buoy motion so that beam alignment and resource allocation can be proactively adjusted.

  	\vspace*{-2mm}
	\subsection{Related Works}\label{S1.1}

	UAV-assisted data collection has been extensively investigated for ocean and remote monitoring networks. Ma \emph{et al.} \cite{ma2020uav} surveyed UAV-assisted ocean monitoring networks and showed that aerial data harvesting can effectively complement sparse maritime infrastructure. Zhang \emph{et al.} \cite{zhang2022energy} further optimized fixed-wing UAV trajectories for maritime data collection from buoys under wind effect, highlighting the importance of propulsion-aware design. Moreover,
	our recent works \cite{Li2025MaritimeEmergency,Li2026ISCPB} studied UAV-enabled integrated sensing and communication (ISAC) for large-scale maritime  monitoring systems, demonstrating the feasibility of ISAC on aerial platforms. 
	However, these works generally assume that sea-surface nodes remain static, thereby overlooking the continuously irregular motion of buoys induced by ocean waves and currents, which may significantly affect sensing and communication performance.
	This motivates the need for mobility-aware beam management techniques for UAV--buoy networks.
	
	In this context, sensing-assisted predictive beamforming has become a prominent approach in mobility systems. The seminal work of Liu \emph{et al.} \cite{Liu2020Radar} exploited ISAC echoes to track vehicular states, showing clear gains over feedback-only beam tracking, but it relied on road-constrained vehicle kinematics. Learning-based predictive beamforming was later explored in vehicular ISAC to bypass explicit channel tracking, improving adaptability at the price of stronger data dependence \cite{Mu2021Deep,Liu2022Learning}. For nonlinear mobility, extended Kalman filtering (EKF)-based trackers were developed for distributed MIMO systems and jitter-aware UAV links \cite{Wang2024Intelligent,Tang2024Jitter}.
	Further studies investigated sensing-assisted beam tracking under different receiver architectures and network scenarios, including hybrid analog--digital arrays, cross-cell beam prediction, and UAV-ISAC systems \cite{Pedraza2025Architectures,Yang2025CrossCell,Zhou2025Temporal}.
	Nevertheless, these studies were developed for tracking vehicles or generic aerial platforms, and generally adopt constant-velocity or geometry-smooth motion models, which are not applicable to ocean buoys experiencing wave-driven oscillations and current-induced drift.
	
	Meanwhile, accurate modeling of sea surface dynamics and sea clutter has long been studied in the maritime radar literature. 
	Sea clutter statistics and modeling techniques were investigated in \cite{Huang2021SeaClutter}, which highlighted the strong dependence of clutter characteristics on sensing geometry and sea state. 
	Compound-Gaussian and heterogeneous clutter models were further analyzed in \cite{Xie2022Compound,Chen2022Heterogeneous}, demonstrating that clutter statistics may significantly deviate from idealized Gaussian assumptions. 
	Recent studies have also shown that sea clutter suppression remains challenging even with advanced signal processing techniques \cite{Zhu2022SeaClutter}. 
	Nevertheless, these studies mainly focus on radar detection problems and rarely consider their impact on communication beam management in UAV-enabled maritime sensing networks.
	
	
	Therefore, existing works still lack a unified framework that jointly considers wave-driven buoy motion, sea clutter, and predictive beam alignment for UAV-enabled maritime data collection. 
	To bridge this gap, this paper develops a sensing-assisted predictive beamforming framework that explicitly incorporates wave-consistent buoy dynamics and clutter-aware sensing models, enabling robust communication between UAVs and drifting ocean buoys.
	
	\vspace*{-2mm}
	\subsection{Contributions}\label{S1.2}

	The main contributions are summarized as follows:
	\begin{itemize}
		\item We establish a clutter-aware sensing-assisted predictive beamforming framework for UAV-enabled ocean monitoring, where a frame-based mission workflow tightly integrates target sensing, EKF-based state refinement, proactive beam prediction, and subsequent uplink collection from buoys under sea-surface dynamics.
		
		\item We develop a physics-grounded yet tractable buoy-motion model that jointly captures wave-induced oscillations and current-induced drift through a correlated-acceleration state-space formulation. In particular, the Singer correlation time is explicitly linked to the dominant wave time scale, making the model well suited to short-horizon beam prediction in maritime environments.
		
		\item We derive the posterior Fisher information matrix (FIM) and the associated posterior Cram\'er--Rao bound (PCRB) for the nonlinear sensing measurement model in the presence of residual sea clutter, and adopt the predicted horizontal-position PCRB as a design-oriented sensing metric. This provides a principled way to quantify how sea-surface dynamics and clutter jointly affect predictive beam alignment.
		
		\item We formulate a worst-buoy joint sensing-power and UAV-position design problem under uplink-rate, UAV-power, and mobility constraints. By combining a Schur-complement reformulation with a lagged successive convex approximation, the resulting subproblem is converted into a convex conic program with tractable complexity scaling in the number of buoys.
		
		\item Extensive simulations demonstrate that the proposed design delivers the most balanced sensing--communication performance among the considered benchmarks and remains robust under denser buoy deployments, harsher sea conditions, and stronger process/measurement noise. In particular, the severe RMSE degradation of the communication-only benchmark verifies the necessity of sensing-assisted state refinement for maritime predictive beamforming. Moreover, compared with a full first-order Taylor expansion method, it incurs only limited performance loss while substantially reducing the optimization time, thereby highlighting its practical advantage in performance--complexity tradeoff for online maritime deployment.
	\end{itemize}

	\begin{figure}[!b]
		\vspace*{-1mm}
		\centering
		\includegraphics[width=\linewidth]{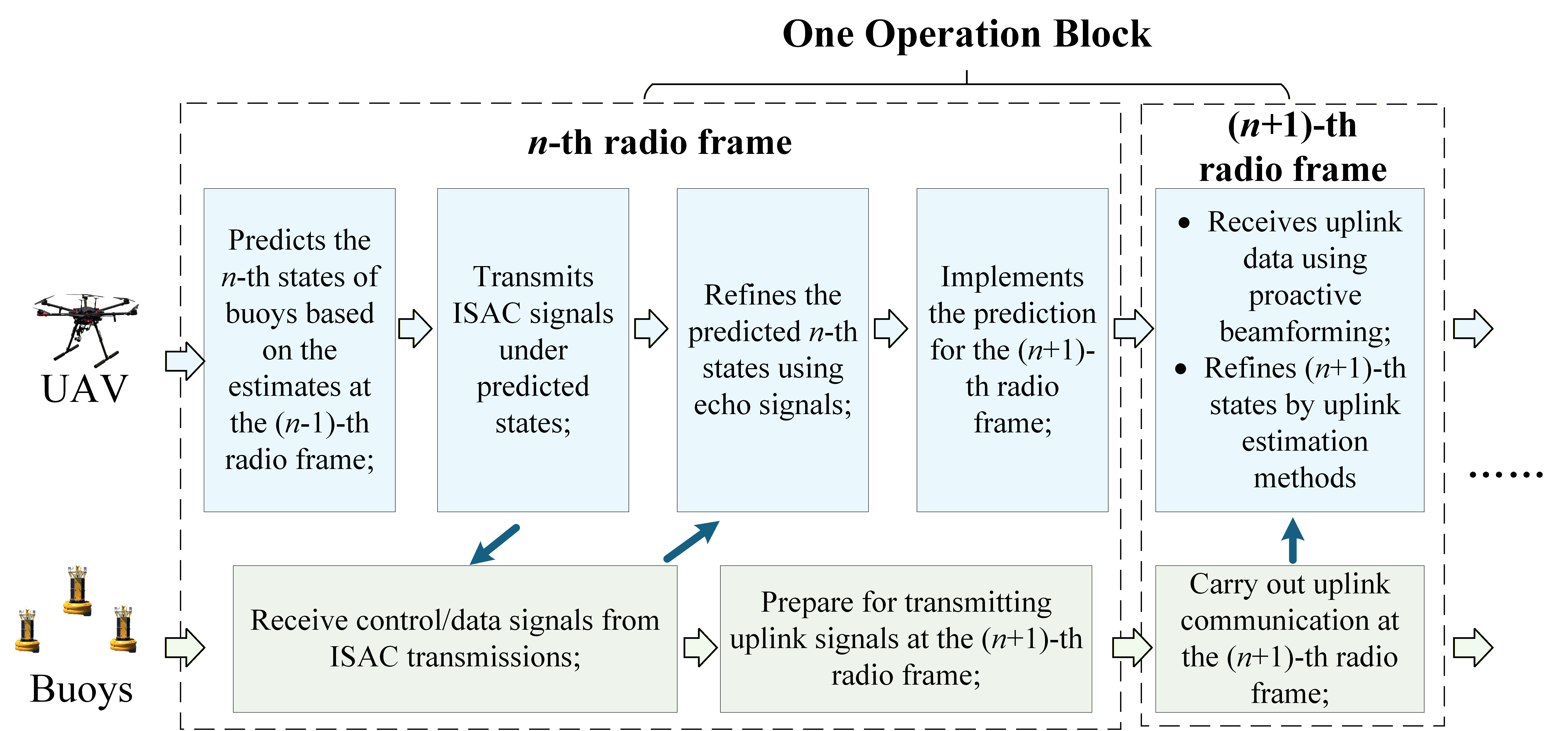}
		\caption{\small Frame structure designed for the UAV mission workflow.}
		\label{figure-UUSW-FS} 
	\end{figure}

\begin{figure*}[!t]\setcounter{equation}{1}
\vspace*{-1mm}
\begin{align}\label{eqrnt}
  \pmb{y}_{n}(t) =& \pmb{z}_n(t) + \sum_{k=1}^{K} \bigg(\underbrace{\beta_{k,n}e^{j2\pi f^{\text{D}}_{k,n}t}\pmb{a}_{\text{rx}}(\theta_{k,n},\phi_{k,n}) \pmb{a }_{\text{tx}}^{\rm H}( \theta_{k,n},\phi_{k,n})\pmb{x}_n(t-\tau_{k,n})}_{\text{Target echo}} \nonumber \\
	& \qquad\qquad\quad + \underbrace{\beta^{\text{clu}}_{k,n}e^{j2\pi f^{\text{D,clu}}_{k,n}t}\pmb{a}_{\text{rx}}(\theta^{\text{clu}}_{k,n},\phi^{\text{clu}}_{k,n}) \pmb{a }_{\text{tx}}^{\rm H}( \theta^{\text{clu}}_{k,n},\phi^{\text{clu}}_{k,n})\pmb{x}_n(t-\tau^{\text{clu}}_{k,n})}_{\text{Sea clutter}}\bigg) ,
\end{align} 
\hrulefill
\vspace*{-1mm}
\end{figure*}	
		
	\vspace*{-1mm}
	\section{System Model}\label{S2}
	
	We consider an ocean monitoring network, where $K$ non-tethered single-antenna buoys are deployed in a designated sea area to monitor the marine ecological environment. To obtain the observation data therein, a rotary-wing UAV equipped with a uniform planar array (UPA) of $\bar{R}=R\times R$ antennas is designed to fly over the buoys and collect data via uplink transmissions. 
	Let $T$ denote the whole mission duration, which is divided into $N$ radio frames of duration $\bar{T}=T/N$. Here, $\bar{T}$ is interpreted as a frame-level scheduling interval. It is  assumed that the UAV--buoy channels are quasi-static only over the much shorter sensing and communication signaling intervals within each radio frame.
	
	\vspace*{-2mm}
	\subsection{UAV Mission Workflow}\label{sec:UMW} 
	
	In practice, the non-negligible sea waves induce continuous motion of the buoys, and any misalignment of the UAV's narrow beam can lead to a severe degradation in communication rate. Therefore, the UAV has to acquire the accurate position information of the buoys, so as to accomplish the reliable uplink communication. However, due to the highly dynamic nature of sea-surface buoys, the buoy's position information obtained at the current radio frame through beam training or tracking cannot be directly used to form the beamforming vector for the next radio frame. As such, beam prediction must be leveraged to proactively estimate the buoy coordinates for the next radio frame, enabling accurate beam alignment.
	
	Accordingly, the overall UAV mission workflow is illustrated in Fig. \ref{figure-UUSW-FS}. Specifically, two consecutive radio frames form one operation block (OB). At the beginning of the mission, the UAV acquires coarse buoy locations via GPS and then heads to the target sea area. Upon arrival, it transmits radar signals to obtain the initial state information of the buoys (i.e., at frame 0). { Then, during each OB, at the $n$-th radio frame, the UAV predicts the buoy states from the estimates of the $(n-1)$-th frame and transmits ISAC signals toward the buoys.
	The ISAC signals deliver the required signaling information and notify the buoys to start observation-data transmission in the subsequent $(n+1)$-th radio frame. Meanwhile, the UAV processes the received ISAC echoes to refine the predicted parameters at frame $n$, and then uses the refined results to generate the proactive beam prediction for the $(n+1)$-th frame. In addition, when receiving the data from buoys, the UAV is able to refine the  $(n+1)$-th states by leveraging uplink pilots or decision-directed data\footnote{The MMSE channel estimation method is adopted during simulations, but for the sake of brevity, the detailed discussion is omitted here.}.}

\vspace*{-2mm}
	\subsection{Signal Model}\label{S2.2}
	
	\subsubsection{Radar Signal Model}
	
	Let the baseband ISAC streams at the $n$-th radio frame be 
	$\bar{\pmb{x}}_n(t)\! =\! [\bar{x}_{1,n}(t),\dots,\bar{x}_{K,n}(t)]^{\text{T}}\! \in\! \mathbb{C}^{K\times 1}$,
	where $\mathbb{E}[\bar{\pmb{x}}_{n}(t)\bar{\pmb{x}}_{n}^{\text{H}}(t)]\! =\! \pmb{I}_{K}, \forall n$. Then, the transmitted signal at the $n$-th radio frame is given by\setcounter{equation}{0}
\begin{align} 
	\pmb{x}_n(t)=\sum_{k=1}^K \sqrt{p_{k, n}} \pmb{f}_{k, n} \bar{x}_{k, n}(t) ,
\end{align}
	where $\pmb{f}_{k,n}\! \in\! \mathbb{C}^{\bar{R}\times 1}$ is the precoding vector and $p_{k,n}$ is the transmit power allocated to sense the $k$-th buoy at the $n$-th radio frame.
	As a result, the received echo signals $\pmb{y}_{n}(t)$ reflected by buoys can be expressed as \eqref{eqrnt} at the top of this page,
	where $f^{\text{D}}_{k,n} (f^{\text{D,clu}}_{k,n})$, $\tau_{k,n}$ ($\tau^{\text{clu}}_{k,n}$), $\theta_{k,n}$ ($\theta_{k,n}^{\text{clu}}$), $\phi_{k,n}$ ($\phi_{k,n}^{\text{clu}}$) and $\beta_{k,n}$ ($\beta^{\text{clu}}_{k,n}$) are the Doppler frequency, time delay, the elevation angle of departure/arrival (E-AoD/AoA), the azimuth angle of departure/arrival (A-AoD/AoA) and reflection coefficient, with respect to the $k$-th buoy (the sea clutter around the $k$-th buoy), respectively, while $\pmb{z}_n(t)\! \sim\! \mathcal{CN}(\pmb{0},N_0\pmb{I}_{\bar{R}})$ is the complex additive white Gaussian noise (AWGN). Both the UAV's transmit and receive array response vectors, i.e., $\pmb{a}_{\text{tx}}(\theta_{k,n},\phi_{k,n})$ and $\pmb{a }_{\text{rx}}( \theta_{k,n},\phi_{k,n})$, can be expressed as\setcounter{equation}{2}
	\begin{align}\label{eqUPV} 
		& \pmb{a}\left(\theta_{k,n},\phi_{k,n}\right) = \frac{1}{R}\left[1,\dots,e^{\frac{-j 2\pi (R-1) d_{\text{a}}}{\lambda}\sin(\theta_{k,n})\cos(\phi_{k,n}) }\right]^{\text{T}} \nonumber \\
		& \hspace*{18mm}\otimes\left[1,\dots,e^{\frac{-j 2\pi (R-1) d_{\text{a}}}{\lambda} \sin(\theta_{k,n})\sin(\phi_{k,n})}\right]^{\text{T}}\!\! ,\!
	\end{align}
	where 
	$d_{\text{a}}$ is the antenna spacing along both the directions of x-axis and y-axis, and $\lambda$ is wavelength. 
	Given $d_{k,n}\! =\! \|\pmb{c}_{\text{U},n}-\pmb{c}_{k,n}\|$, with  $\pmb{c}_{k,n}\! =\! [c^x_{k,n},c^y_{k,n},c^z_{k,n}]^{\text{T}}$ and $\pmb{c}_{\text{U},n}\! =\! [c^x_{\text{U},n},c^y_{\text{U},n},c^z_{\text{U},n}]^{\text{T}}$ denoting the 3D-coordinates of buoy $k$ and the UAV, respectively, we have
	\begin{align} 
		\hspace*{-2mm}\sin(\theta_{k,n})\cos(\phi_{k,n})\! =\! \frac{\Delta_{k,n}^x}{d_{k,n}}, \ 
		\sin(\theta_{k,n})\sin(\phi_{k,n})\! =\! \frac{\Delta_{k,n}^y}{d_{k,n}} ,\!
	\end{align}
	where $\Delta_{k,n}^{x(y)}=c^{x(y)}_{\text{U},n}-c^{x(y)}_{k,n}$.
	Then, $\pmb{a}\left(\theta_{k,n},\phi_{k,n}\right)$ can be substituted with $\pmb{a}\left(\pmb{c}_{k,n},\pmb{c}_{\text{U},n}\right)$, which is given by $\pmb{a}(\pmb{c}_{k,n},\pmb{c}_{\text{U},n})=\frac{1}{R}\pmb{a}^{x}(\pmb{c}_{k,n},\pmb{c}_{\text{U},n})\otimes \pmb{a}^{y}(\pmb{c}_{k,n},\pmb{c}_{\text{U},n})$,
	where
	\begin{align} 
	\pmb{a}^{x(y)}(\pmb{c}_{k,n},\pmb{c}_{\text{U},n})=\left[1,e^{-j\mu_{k,n}^{x(y)}},\dots,e^{-j(R-1)\mu_{k,n}^{x(y)}}\right]^{\text{T}},
	\end{align} 
	and
	$\mu_{k,n}^{x(y)}=\frac{-2\pi  d_{\text{a}}\Delta_{k,n}^{x(y)}}{\lambda d_{k,n}}$. 
	
	Since the channel between the UAV and each buoy is dominated by the LoS path, $\pmb{f}_{k,n}$ is readily designed based on the transmit array response vector. However, due to the continuous movement of buoys, the parameters estimated at the previous radio frame are no longer valid at the subsequent radio frame. Therefore, we have
	 $\pmb{f}_{k,n}\! =\! \pmb{a}_{\text{tx}}(\hat{\pmb{c}}_{k,n|n-1},\pmb{c}_{\text{U},n}), \forall k={1,\dots,K}$, where $\hat{\pmb{c}}_{k,n|n-1}$ is the one-step predicted coordinate of the $k$-th buoy at the $n$-th radio frame. 
	
	\begin{figure*}[!t]\setcounter{equation}{11}
	\vspace*{-1mm}
		\begin{align}\label{eqrknt} 
			\pmb{y}_{k,n}(t) =& \pmb{z}_{k,n}(t) + \sqrt{p_{k,n}}
			e^{j2\pi f^{\text{D}}_{k,n}t}\bigg(\beta_{k,n}\pmb{a}_{\text{rx}}(\pmb{c}_{k,n},\pmb{c}_{\text{U},n}) \pmb{a }_{\text{tx}}^{\rm H}(\pmb{c}_{k,n},\pmb{c}_{\text{U},n}) \nonumber \\
			& \qquad\qquad\qquad\qquad\qquad +\rho_{k,n}\beta^{\text{clu}}_{k,n}\pmb{a}_{\text{rx}}(\hat{\pmb{c}}_{k,n|n-1},\pmb{c}_{\text{U},n}) \pmb{a }_{\text{tx}}^{\rm H}(\hat{\pmb{c}}_{k,n|n-1},\pmb{c}_{\text{U},n})\bigg)\pmb{f}_{k,n}\bar{x}_{k,n}(t-\tau_{k,n}) .
		\end{align} 
		\hrulefill
		\vspace*{-2mm}
	\end{figure*}
	
	\begin{remark} 
		The received echo signals consist of reflections from the buoys as well as sea waves in their vicinity. With a massive MIMO array deployed at the UAV, the highly directional transmit beam significantly confines the illuminated area, such that the resulting sea clutter predominantly originates from the transmit beam direction. Hence, we have\setcounter{equation}{5}
	\begin{align}\label{eq:thetaclu} 
		\theta^{\text{clu}}_{k,n}=\hat{\theta}_{k,n|n-1},\  \phi^{\text{clu}}_{k,n}=\hat{\phi}_{k,n|n-1}, \  \tau^{\text{clu}}_{k,n}=\tau_{k,n}.
	\end{align}  
		In addition, it is assumed that the Doppler-domain clutter suppression is applied to mitigate part of sea clutter components (e.g., Bragg peaks) \cite{ward2006sea}. Nevertheless, due to Doppler spreading and finite-time processing, another part of clutter may leak into the target Doppler bin. This residual component is therefore modeled as an additional low-power interference term sharing the same Doppler bin as the target. Accordingly, the Doppler term of sea clutter in \eqref{eqrnt} can be modeled as $e^{j2\pi f^{\text{D,clu}}_{k,n}t}\approx \rho_{k,n} \cdot e^{j2\pi f^{\text{D}}_{k,n}t}$, where $0< \rho_{k,n} <1$ accounts for effectiveness of Doppler-domain clutter suppression.
	\end{remark}
	
	\subsubsection{Reflection Coefficient of Buoy and Sea Clutter}
	In \eqref{eqrnt}, $\beta_{k,n}$ ($\beta^{\text{clu}}_{k,n}$)  is determined by the signal propagation distance and the radar cross section (RCS) of buoy (sea clutter). Specifically, for $\beta_{k,n}$, we have $\beta_{k,n} = \frac{\chi_{k,n}}{2d_{k,n}}$, where $\chi_{k,n}$ is the complex RCS of buoy $k$ at the $n$-th radio frame. As for $\beta^{\text{clu}}_{k,n}$, the RCS of sea clutter depends not only on the backscattering coefficient $\bar{\chi}_{k,n}^{\text{clu}}$ but also the area of sea clutter patch $A_{k,n}$, i.e., $\chi_{k,n}^{\text{clu}}=\bar{\chi}_{k,n}^{\text{clu}}A_{k,n}$. In this work, $\bar{\chi}_{k,n}^{\text{clu}}$ is determined by the widely-used Morchin model, which is given by \cite{clarke1985airborne} 
	\begin{align}\label{eqMm} 
		\bar{\chi}_{k,n}^{\text{clu}} =& \frac{4\times 10^{0.6\left(\kappa_{\text{s}}+1\right)-7}}{\lambda}{\chi}_{k,n}^{0}\sin\varphi_{k,n} \nonumber \\
		& + {\Gamma_\text{s}e^{-\tan^2\left(\pi/2-\varphi_{k,n}\right)\Gamma_\text{s}}} ,
	\end{align} 
	where the integer $0\! \leq\! \kappa_{\text{s}}\! \leq\! 9$ defines the sea state, $\varphi_{k,n}$ is the grazing angle, $\Gamma_\text{s}=\cot^2\left(2.44(\kappa_{\text{s}}+1)^{1.08}/57.29\right)$, and ${\chi}_{k,n}^{0}$ is defined by
	\begin{equation}\label{eqMm1} 
		{\chi}_{k,n}^{0} = \begin{cases}
			\left(\frac{\varphi_{k,n}}{\varphi_{0}}\right)^{1.9}, \ &\varphi_{k,n}\leq\varphi_{0} , \\
			1,\  &\varphi_{k,n} > \varphi_{0} ,\ 
		\end{cases}
	\end{equation}
	with $\varphi_{0}\! =\! \text{arcsin}\left(\lambda/(0.1\pi+0.184\pi\kappa_{\text{s}})\right)$. 
	It is worth noting that under typical UAV flight geometries and sea conditions, the inequality $\varphi_{k,n} > \varphi_{0}$ is satisfied with high probability. Hence, for analytical simplification, we approximate ${\chi}_{k,n}^{0}\approx 1$. Then, \eqref{eqMm} can be rewritten as
	\begin{align}\label{eqchiclu} 
		\bar{\chi}_{k,n}^{\text{clu}} =\frac{4\times 10^{0.6\left(\kappa_{\text{s}}+1\right)-7}}{\lambda}\frac{c^z_{\text{U},n}}{d_{k,n}}+\Gamma_\text{s}e^{-\Gamma_\text{s}\big(\frac{\bar{d}_{k,n}}{c^z_{\text{U},n}}\big)^2} ,
	\end{align}
	where $\bar{d}_{k,n}=\sqrt{(\Delta_{k,n}^x)^2+(\Delta_{k,n}^y)^2}$. The area of sea clutter patch can be derived as $A^{\text{sc}}_{k,n}\! =\! \frac{c_0\, d_{k,n}\phi_{\text{3dB}}}{2B\cos\left(\varphi_{k,n}\right)}$, where $c_0$ and $B$ are light speed and bandwidth, respectively, and $\phi_{\text{3dB}}$ denotes the half-power beamwidth in the azimuth plane, which is given by $\phi_{\text{3dB}}\! =\!\frac{0.886\lambda}{R\cdot d_{\text{a}}}$ \cite{balanis2015antenna}. Then, we have $A^{\text{sc}}_{k,n}=\frac{0.886c_0\cdot d_{k,n}^2}{BR\bar{d}_{k,n}}$ under the assumption of $d_{\text{a}}=\lambda/2$, and 
	\begin{align}\label{eqbetaclu} 
		\beta^{\text{clu}}_{k,n} = \Xi_1 \frac{c^z_{\text{U},n}}{\bar{d}_{k,n}}+\Xi_2\frac{d_{k,n}}{\bar{d}_{k,n}}e^{-\Gamma_\text{s}\big(\frac{\bar{d}_{k,n}}{c^z_{\text{U},n}}\big)^2} ,
	\end{align}
	where 
	\begin{align} 
		\Xi_1 = \frac{1.772 c_0\times 10^{0.6\left(\kappa_{\text{s}}+1\right)-7}}{B\cdot R\lambda}, \ \Xi_2 =\frac{0.443c_0 \cdot \Gamma_\text{s}}{B \cdot R}.
	\end{align}
	
	\begin{remark} 
		In this work, the RCS of buoy $k$ is assumed to be time-invariant over the considered beam update interval, i.e., $\chi_{k,n}\approx \chi_{k,n-1}, \forall k$\footnote{This approximation is suitable for the short-horizon predictive beamforming design studied in this paper. More detailed RCS fluctuations caused by buoy pitch/roll motions or Swerling-type scattering are not explicitly modeled, and their robust treatment is left for future work.}. This assumption is justified since the buoy is modeled as a rigid structure without rapid orientation changes, and thus its scattering property remains approximately constant within a short prediction horizon. In contrast, the RCS of the sea clutter is inherently time-varying. In particular, according to \eqref{eqchiclu}, ${\chi}_{k,n}^{\text{clu}}$  is not constant but varies with the instantaneous coordinates of the UAV and the buoy due to the changing propagation geometry.
	\end{remark}
	
	\subsubsection{Radar Measurement Model}
	As the UAV employs a massive MIMO array, the pencil-like beam can largely mitigate the multi-target interference at the receiver \cite{dong2022sensing}. Also, in maritime scenarios, compared with terrestrial targets, buoys are typically sparsely distributed, resulting in a more pronounced separation in the angular domain. Hence, given \eqref{eq:thetaclu}, the echo signal of buoy $k$ can be expressed as \eqref{eqrknt} at the top of this page.
	Due to the fact that the sea area within the beam footprint is limited and the reflection coefficient of sea is generally smaller than that of buoy \cite{ward2006sea}, the energy of buoy echo dominates in \eqref{eqrknt}. In this regard, the matched-filtering method can be applied to estimate $f^{\text{D}}_{k,n}$ and $\tau_{k,n}$ for buoy $k$, which can be expressed as\setcounter{equation}{12}
	\begin{align} 
		\{\hat{f}^{\text{D}}_{k,n},\hat{\tau}_{k,n}\}\! =\!	\arg\max\limits_{f^{\text{D}},\tau}\bigg|\! \int_0^{\bar{T}}\!\! \pmb{y}_{k,n}(t)\bar{x}_{k,n}^{*}(t-\tau)e^{-j2\pi f^{\text{D}}t}dt\bigg|^2\!\! .\!
	\end{align}
	Then, with these estimates, the matched-filtering result can be derived as
	\begin{align}\label{eqint0} 
		&\int_0^{\bar{T}}\pmb{y}_{k,n}(t)\bar{x}_{k,n}^{*}(t-\hat{\tau}_{k,n})e^{-j2\pi \hat{f}^{\text{D}}_{k,n}t}dt \nonumber \\
		& =
		\sqrt{p_{k,n}}M_{\text{f}}\bigg(\beta_{k,n}\pmb{a}_{\text{rx}}(\pmb{c}_{k,n},\pmb{c}_{\text{U},n}) \pmb{a }_{\text{tx}}^{\rm H}( \pmb{c}_{k,n},\pmb{c}_{\text{U},n}) + \rho_{k,n} \beta^{\text{clu}}_{k,n} \nonumber \\
		&\times \pmb{a}_{\text{rx}}(\hat{\pmb{c}}_{k,n|n-1},\pmb{c}_{\text{U},n}) \pmb{a }_{\text{tx}}^{\rm H}( \hat{\pmb{c}}_{k,n|n-1},\pmb{c}_{\text{U},n})\bigg)\pmb{f}_{k,n}\! +\! \pmb{z}_{1,k,n} ,
	\end{align}
	where $\pmb{z}_{1,k,n}\! =\! \int_0^{\bar{T}}\pmb{z}_{k,n}(t)\bar{x}_{k,n}^{*}(t-\hat{\tau}_{k,n})e^{-j2\pi \hat{f}^{\text{D}}_{k,n}t}dt$, and
	$M_{\text{f}}\! =\! \int_0^{\bar{T}}\bar{x}_{k,n}(t-\tau_{k,n})\bar{x}_{k,n}^{*}(t-\hat{\tau}_{k,n})e^{j2\pi ({f}^{\text{D}}_{k,n}-\hat{f}^{\text{D}}_{k,n})t}dt$ is the matched-filtering gain. After the matched-filtering and normalization processes, the measurement model for $\pmb{c}_{k,n}$ and $\beta_{k,n}$ can be obtained as
	\begin{align}\label{eqykn} 
		\pmb{r}_{k,n} =& 
		\beta_{k,n}\varpi_{k,n}^{\text{tx}}\pmb{a}_{\text{rx}}(\pmb{c}_{k,n},\pmb{c}_{\text{U},n})+
		\rho_{k,n}\beta^{\text{clu}}_{k,n} \nonumber \\
		& \times \pmb{a}_{\text{rx}}(\hat{\pmb{c}}_{k,n|n-1},\pmb{c}_{\text{U},n})+\tilde{\pmb{z}}_{1,k,n} ,
	\end{align}
	where
	$\varpi_{k,n}^{\text{tx}}\! =\! \pmb{a }_{\text{tx}}^{\rm H}(\pmb{c}_{k,n},\pmb{c}_{\text{U},n})\pmb{a}_{\text{tx}}(\hat{\pmb{c}}_{k,n|n-1},\pmb{c}_{\text{U},n})$. Since $\pmb{z}_{1,k,n}$ is obtained as the time integral of the product of $\pmb{z}_{k,n}(t)$ and $\bar{x}_{k,n}(t)$, the normalized $\tilde{\pmb{z}}_{1,k,n}$ has zero mean and a variance of $\sigma_{1,k,n}^2$, where $\sigma_{1,k,n}^2={c_1^2 N_0}/({{p_{k,n}}M_{\text{f}}^2})$ and $c_1$ is constant related to the signal design and system configuration.
	
	Also, the measurement model for $d_{k,n}$ and $v_{k,n}$ can be obtained as
	\begin{align}\label{eqfmea} 
		{f}^{\text{D}}_{k,n} = \frac{2}{\lambda}\pmb{e}_{k,n}^{\text{T}}\pmb{v}_{k,n}+z_{2,k,n}, \ \  
		{\tau}_{k,n} = \frac{2d_{k,n}}{c_0} + z_{3,k,n} ,
	\end{align}
	where $\pmb{e}_{k,n}=\frac{\pmb{c}_{\text{U},n}-\pmb{c}_{k,n}}{d_{k,n}}$ denotes the unit direction vector pointing from the UAV to the $k$-th buoy, $\pmb{v}_{k,n}=[v^x_{k,n},v^y_{k,n},0]$ denotes the horizontal velocity of buoy $k$. According to \cite{kay1993fundamentals}, the variances of the measurement noises are inversely proportional to the receive signal-clutter-noise ratio (SCNR) of \eqref{eqint0}. Hence, similar to $\sigma_{1,k,n}^2$, we can obtain
	\begin{align} 
		\sigma_{i,k,n}^2 = \frac{c_i^2(N_0+p_{k,n}M_{\text{f}}^2\rho_{k,n}^2|\beta^{\text{clu}}_{k,n}|^2)}{p_{k,n}M_{\text{f}}^2 |\beta_{k,n}|^2|\varpi_{k,n}^{\text{tx}}|^2}, \ i=2,3 ,
	\end{align}
	where $c_i$ ($i\! =\! 2,3$) are constants determined by the signal design and system parameters. In the considered beam-tracking regime, the EKF-based one-step prediction (OSP) is used to keep the angular mismatch within the mainlobe of antenna beam over consecutive frames, and hence we can assume that $\varpi_{k,n}^{\text{tx}}\! \approx\! 1$.
	
	\subsubsection{Uplink Communication Model}
	In the proposed data-collection scenario, we primarily consider the more critical uplink, by which the buoys transmit their observation data to the UAV, while the downlink from the UAV to the buoys (e.g., control signaling) is not explicitly modeled. {Also, we assume that the UAV adjusts its position only once within one OB, due to the fact that the positions of buoys change slightly over two consecutive frames. Hence, it is not necessary for the UAV to frequently change its position, for the sake of both computational complexity and energy consumption. Accordingly, at the $(n+1)$-th radio frame, the received signal at the UAV can be expressed as
		\begin{align}\label{eqtirknt} 
			\tilde{y}_{k,n+1}(t) = &
			\sqrt{P_{\text{b}}}\alpha_{k,n+1} e^{j2\pi f^{\text{D}}_{k,n+1}t}\pmb{w}_{k,n+1}^{\text{H}}\pmb{a}_{\text{rx}}(\pmb{c}_{k,n+1},\pmb{c}_{\text{U},n}) \nonumber \\
			& \times \tilde{x}_{k,n+1}(t)+ z_{\text{c}}(t) ,
		\end{align} 
		where $P_{\text{b}}$ is the power budget of buoy, $\alpha_{k,n+1}=\tilde{\alpha}d_{k,n+1}^{-1}e^{j\frac{2\pi}{\lambda}d_{k,n+1}}$ denotes the uplink channel coefficient with $\tilde{\alpha}$ being the channel power gain at the reference distance of 1m, $\tilde{x}_{k,n+1}(t)$ is the transmitted signal of buoy $k$,} and $z_{\text{c}}(t)\sim\mathcal{CN}(0,N_0)$ denotes the AWGN. Note that as the Doppler frequency shift $f^{\text{D}}_{k,n+1}$ can be estimated at the UAV, it will be largely compensated at the receiver. Moreover, with the aid of narrow receive beam generated by UAV's massive antenna array, the inter-buoy interference is neglected.
	
	The design of the combining vector $\pmb{w}_{k,n+1}$ is highly related to the receive array response vector. As mentioned in Section \ref{sec:UMW}, the UAV leverages the predicted position information (i.e., $\hat{\pmb{c}}_{k,n|n-1}$) to transmit ISAC signals. Then, relying on the echo signals, the angle information is further revised to output the next prediction (i.e., $\hat{\pmb{c}}_{k,n+1|n}$), which is used to compute $\pmb{w}_{k,n+1}$. Hence, at the $(n+1)$-th radio frame, the combining vector is derived as $\pmb{w}_{k,n+1}=\pmb{a}_{\text{rx}}(\hat{\pmb{c}}_{k,n+1|n},\pmb{c}_{\text{U},n})$.
	
	\vspace*{-2mm}
	\subsection{State Evolution Model}
	
	To enhance the uplink communication rate, it is necessary to predict and track the short-term motion trajectory of the buoy. The horizontal motion of a surface buoy is primarily driven by two physical factors, ocean currents and sea waves. 
	The ocean current is typically characterized by a relatively slow spatial and temporal variation and therefore, within a short-term prediction horizon, can be reasonably approximated as a constant background drift.  Let $v_{\text{c}}^{x(y)}$ denote the constant ocean-current-induced drift velocity.
	In contrast, sea waves introduce high-frequency oscillatory disturbances and stochastic excitation, leading to random and temporally correlated motion components. Let $\bar{v}_{k,n}^{x(y)}(t)$ denote the wave-induced velocity at time instant $t$.
	As a result, the buoy velocity can be viewed as the superposition of a deterministic current-induced drift and a wave-induced random fluctuation, i.e., ${v}_{k,n}^{x(y)}(t)=\bar{v}_{k,n}^{x(y)}(t)+v_{\text{c}}^{x(y)}$, while the acceleration dynamics are mainly governed by wave-driven environmental excitation.
	
	In the continuous-time domain, the kinematic equations are expressed as 
	\begin{align}\label{eqdotc} 
		\dot{c}^{x(y)}_{k,n}(t)=v_{k,n}^{x(y)}(t), \ \dot{v}^{x(y)}_{k,n}(t)=a_{k,n}^{x(y)}(t),
	\end{align}
	where $\dot{c}^{x(y)}_{k,n}(t)$ and $\dot{v}^{x(y)}_{k,n}(t)$ denote the time derivative of the position and velocity along the direction of $x-$axis ($y-$axis), respectively.
	The short-term dynamic variation caused by sea waves is mainly reflected in the acceleration behavior. To capture its temporal correlation while maintaining 
	a tractable state-space formulation, 
	the Singer model is adopted to characterize 
	the buoy acceleration as a first-order Gaussian Markov process \cite{singer2007estimating}, and it is assumed that the accelerations in $x$ and $y$ axes are modeled as independent Singer processes with the same characteristics. Then, $a^{x(y)}_{k,n}(t)$ can be modeled as 
	\begin{align}\label{eq:daknt} 
		\dot{a}_{k,n}^{x(y)}(t) = -\varsigma{a}_{k,n}^{x(y)}(t) + z_{\text{a}}^{x(y)}(t),
	\end{align}
	where $\dot{a}_{k,n}^{x(y)}(t)$ represents the time derivative of the acceleration, $\varsigma>0$  is the reciprocal of the acceleration time constant. $z_{\text{a}}^{x}(t)$ and $z_{\text{a}}^{y}(t)$ are mutually independent zero-mean white Gaussian processes satisfying $z_{\text{a}}^{x(y)}(t)\sim\mathcal{N}(0,2\varsigma\sigma_a^2)$. The corresponding autocorrelation function of the acceleration is given by
	\begin{equation} 
		R_a(\tau) = \sigma_a^2 e^{-\varsigma |\tau|},
	\end{equation}
	where $\tau$ denotes the time lag between two observations. 
	
	\begin{remark} 
		For a surface buoy, the dominant short-term excitation is wave-induced, and the buoy’s horizontal acceleration varies on the same time scale as the dominant wave component. Let the dominant wave period be $T_{\text{p}}$ and the corresponding angular frequency be $\omega_{\text{p}}\! =\! \frac{2\pi}{T_{\text{p}}}$. Then, it is physically reasonable to select the Singer time constant to be on the order of the dominant wave time scale, i.e., $\varsigma\! \approx\! \omega_{\text{p}}$, which endows the stochastic acceleration with a temporal memory consistent with the dominant wave-induced oscillations. 
	\end{remark}
	
	Given \eqref{eqdotc} and \eqref{eq:daknt}, we have 
	\begin{align}\label{eq:dotskn} 
		\dot{\pmb{s}}_{k,n}(t) = &\text{Blkdiag}(\pmb{\Phi}_{\text{s}},\pmb{\Phi}_{\text{s}})\pmb{s}_{k,n}(t) + \pmb{b}_{\text{c}} \nonumber \\
		& + \text{Blkdiag}(\pmb{e}_{\text{s}},\pmb{e}_{\text{s}})\pmb z_{\text{a}}(t),
	\end{align}
	where $\pmb{\Phi}_{\text{s}}\! =\! [0,1,0;0,0,1;0, 0, -\varsigma]$,
	$\pmb{s}_{k,n}(t)\! =\! [\pmb{s}^x_{k,n}(t);\pmb{s}^y_{k,n}(t)]$ with $\pmb{s}^{x(y)}_{k,n}(t)\! =\! [c^{x(y)}_{k,n}(t), \bar{v}_{k,n}^{x(y)}(t), a_{k,n}^{x(y)}(t)]^{\text{T}}$, 
	$\pmb{b}_{\text{c}}\! =\! [v_{\text{c}}^x,0,0,v_{\text{c}}^y,0,0]^{\text{T}}$, 
  $\pmb{e}_{\text{s}}\! =\! [0,0,1]^{\text{T}}$, and $\pmb z_{\text{a}}(t)\! =\! [z_{\text{a}}^{x}(t), z_{\text{a}}^{y}(t)]^{\text T}$.
	Then, by discretizing the continuous-time dynamics in \eqref{eq:dotskn} with the sampling interval $\bar{T}$, the state evolution can be expressed in discrete time as
	\begin{align}\label{eqskn1} 
		\pmb{s}_{k,n+1} = \text{Blkdiag}\Big(\bar{\pmb{\Phi}}_{\text{s}},\bar{\pmb{\Phi}}_{\text{s}}\Big)\pmb{s}_{k,n}+ \bar{T}\pmb{b}_{\text{c}}+\pmb{u}_{\text{s}},
	\end{align}
	where $\bar{\pmb{\Phi}}_{\text{s}}\! =\! [1,\bar{T},\phi_{1};0,1,\phi_{2};0,0,e^{-\varsigma\bar{T}}]$ with 
 \begin{align} 
		\phi_{1}=\frac{\varsigma\bar{T}-1+e^{-\varsigma\bar{T}}}{\varsigma^2}, \ \ \phi_{2}=\frac{1-e^{-\varsigma\bar{T}}}{\varsigma} ,
	\end{align}	
	$\pmb{s}_{k,n}\! =\! [\pmb{s}^x_{k,n};\pmb{s}^y_{k,n}]$ with $\pmb{s}^{x(y)}_{k,n}=[c^{x(y)}_{k,n}, \bar{v}_{k,n}^{x(y)}, a_{k,n}^{x(y)}]^{\text{T}}$, and 
  $\pmb{u}_{\text{s}}\! =\! [\bar{\pmb{u}}_{\text{s}}^x;\bar{\pmb{u}}_{\text{s}}^y]\! \sim\! \mathcal{N}\big(\pmb{0},\pmb{Q}_{\text{s}}\big)$ is the stacked discrete-time process noise, in which $\bar{\pmb{u}}_{\text{s}}^{x}$ and $\bar{\pmb{u}}_{\text{s}}^{y}$ are independent and identically distributed with $\bar{\pmb{u}}_{\text{s}}^{x(y)}\! \sim\! \mathcal{N}\big(\pmb{0},\bar{\pmb{Q}}_{\text{s}}\big)$ and $\pmb{Q}_{\text{s}}\! =\! \text{Blkdiag}(\bar{\pmb{Q}}_{\text{s}},\bar{\pmb{Q}}_{\text{s}})$. The derivation of $\bar{\pmb{Q}}_{\text{s}}$ is given in Appendix \ref{App:DC}.
	
	\vspace*{-1mm}
	\section{Performance Evaluation and Optimization Formulation}\label{S3}
	
	\vspace*{-1mm}
	\subsection{Radar-Related Evaluation}\label{S3.1}
	
	In the proposed scenario, the UAV transmits ISAC signals towards the buoys on the sea surface, and carries out beam prediction relying on echoes. Therefore, the effectiveness and accuracy of beam prediction serve as the key indicators of radar performance. In the following, we will first present the implementation of beam prediction based on the EKF, and then derive the corresponding PCRB.
	
	\subsubsection{Beam Prediction Using EKF}
	The state evolution and measurement models can be rewritten in a compact version, given by 
	\begin{align}\label{eqekf} 
		\hspace*{-2mm}\begin{cases}
			\! \text{State Evolution Model:} \  \pmb{s}_{k,n} = \pmb{G}\pmb{s}_{k,n-1}+\bar{T}\pmb{b}_{\text{c}}+\pmb{u}_{\text{s}}, \!\!\! \\
			\! \text{Measurement Model:} \ 
			\pmb{m}_{k,n} = h(\pmb{s}_{k,n})+\pmb{u}_{\text{m}}, \!\!\! 
		\end{cases}\!\!
	\end{align}
	where $\pmb{G}=\text{Blkdiag}(\bar{\pmb{\Phi}}_{\text{s}},\bar{\pmb{\Phi}}_{\text{s}})$ and $h(\cdot)$ represents the function of state measurement, $\pmb{m}_{k,n}=[\pmb{r}_{k,n}^{\text{T}}, {f}^{\text{D}}_{k,n}, {\tau}_{k,n}]^{\text{T}}$ denotes the measurement vector, and $\pmb{u}_{\text{m}}=[\tilde{\pmb{z}}_{1,k,n}^{\text{T}},z_{2,k,n},z_{3,k,n}] \sim \mathcal{CN}(\pmb{0},\pmb{Q}_{\text{m}})$ is the stacked measurement noise with $\pmb{Q}_{\text{m}}=\text{Blkdiag}(\sigma_{1,k,n}^2\pmb{I}_{\bar{R}},2\sigma_{2,k,n}^2,2\sigma_{3,k,n}^2)$\footnote{ The factor $2$ in the Doppler/delay blocks ensures that the resulting quadratic form is equivalent to
		the real-valued Gaussian error model with variances $\sigma^2_{2,k,n}$ and $\sigma^2_{3,k,n}$.}. It is worth noting that, as $h(\cdot)$ is non-linear, we resort to the EKF method, and the linearization of measurement model is given by
	\begin{align} 
		\pmb{m}_{k,n} &\approx h(\hat{\pmb{s}}_{k,n|n-1})+\pmb{H}_{k,n}({\pmb{s}}_{k,n}-\hat{\pmb{s}}_{k,n|n-1})+\pmb{u}_{\text{m}} ,
	\end{align}
	where 
	$\pmb{H}_{k,n}=\frac{\partial h}{\partial \pmb{s}}|_{\pmb{s}=\hat{\pmb{s}}_{k,n|n-1}}$ is Jacobian matrix of 
	$h(\pmb{s}_{k,n})$. Following the classic procedure of EKF \cite{ribeiro2004kalman}, the implementation of beam prediction can be summarized as follows.
	\begin{enumerate}[label=(\roman*), leftmargin=*, nosep]
		
		\item \textbf{Prediction}
		\begin{itemize}[leftmargin=1em]
			\item State prediction:
			$\hat{\pmb{s}}_{k,n|n-1}=\pmb{G}\hat{\pmb{s}}_{k,n-1}+\bar{T}\pmb{b}_{\text{c}}$.
			
			\item Mean square error (MSE) matrix prediction:\\
			$\pmb{P}_{k,n|n-1}
			=\pmb{G}
			\pmb{P}_{k,n-1}
			\pmb{G}^{\text{H}}
			+\pmb{Q}_{\text{s}}$  .
		\end{itemize}
		
		\item \textbf{Kalman Gain Computation}
		\begin{itemize}[leftmargin=1em]
			\item
			$\pmb{K}_{k,n}=
			\pmb{P}_{k,n|n-1}
			\pmb{H}_{k,n}^{\text{H}}
			(
			\pmb{H}_{k,n}
			\pmb{P}_{k,n|n-1}
			\pmb{H}_{k,n}^{\text{H}}
			+\pmb{Q}_{\text{m}}
			)^{-1}$.
		\end{itemize}
		
		\item \textbf{Update}
		\begin{itemize}[leftmargin=1em]
			
			\item State estimation update:\\
			$\hat{\pmb{s}}_{k,n}
			=
			\hat{\pmb{s}}_{k,n|n-1}
			+
			\pmb{K}_{k,n}
			\left(
			\pmb{m}_{k,n}
			-
			h(\hat{\pmb{s}}_{k,n|n-1})
			\right)$.
			
			\item MSE matrix update:\\
			$\pmb{P}_{k,n}
			=
			(\pmb{I}-\pmb{K}_{k,n}\pmb{H}_{k,n})
			\pmb{P}_{k,n|n-1}$.
			
		\end{itemize}
		
	\end{enumerate}
	
	\subsubsection{Derivation of PCRB for Beam Prediction}
	
	To characterize the fundamental performance limit of beam prediction based on the ISAC echoes, we derive the PCRB for the buoy state estimation, which incorporates both the measurement information and the state evolution prior. Following the Bayesian estimation framework, the joint probability density function (PDF) of $\pmb{s}_{k,n}$ and $\pmb{m}_{k,n}$ can be expressed via Bayes’ theorem as
	\begin{equation} 
		p(\pmb{s}_{k,n}, \pmb{m}_{k,n}) 
		= p(\pmb{m}_{k,n}| \pmb{s}_{k,n})\, p(\pmb{s}_{k,n}). 
	\end{equation}
	
	With the measurement noise $\pmb{u}_m$, the conditional likelihood admits the complex Gaussian form
	\begin{align} 
		&p(\pmb{m}_{k,n} | \pmb{s}_{k,n})
		=\nonumber \\
		& \hspace*{3mm}\frac{\exp\!\left(
			-
			(\pmb{m}_{k,n} - {h}(\pmb{s}_{k,n}))^{\text H}
			\pmb{Q}_{\text{m}}^{-1}
			(\pmb{m}_{k,n} - {h}(\pmb{s}_{k,n}))
			\right)}{\pi^{\bar{R}+2}\det(\pmb{Q}_{\text{m}})} .
	\end{align}
	
	The prior $p(\pmb{s}_{k,n})$ is determined by the state evolution model and the state information at the $n-1$ radio frame. In particular, 
	given $\pmb{s}_{k,n-1} \sim \mathcal{N}(\hat{\pmb{s}}_{k,n-1}, \pmb{P}_{k,n-1})$, we can derive
	$\pmb{s}_{k,n} \sim \mathcal{N}\big(\hat{\pmb{s}}_{k,n|n-1}, \pmb{P}_{k,n|n-1}\big)$.
	Hence, the prior PDF is\footnote{Note that the kinematic state $\pmb s_{k,n}$ is real-valued, 
		and hence the prior distribution is modeled as a real Gaussian as in (\ref{eq:prior_real}).}
	\begin{align}\label{eq:prior_real} 
		&p(\pmb{s}_{k,n})= \nonumber \\
		&\frac{\exp\big(-\frac{1}{2}
			(\pmb{s}_{k,n}-\hat{\pmb{s}}_{k,n|n-1})^{\text{T}}
			\pmb{P}_{k,n|n-1}^{-1}
			(\pmb{s}_{k,n}-\hat{\pmb{s}}_{k,n|n-1}
			)\big)}{(2\pi)^3\sqrt{ \det(\pmb{P}_{k,n|n-1})}}.
	\end{align}
	
	Consequently, the posterior FIM for $\pmb{s}_{k,n}$ can be derived as 
	\begin{align}\label{eq:FIM_realparam} 
		\pmb{J}_{k,n}
		&=
		-
		\mathbb{E}
		\left[
		\frac{\partial^2\ln p(\pmb{s}_{k,n}, \pmb{m}_{k,n})}{\partial \pmb{s}_{k,n}^2}
		\right] \nonumber \\
		&= 2\Re \left\{(\frac{\partial h}{\partial \pmb{s}_{k,n}})^{\text{H}}\pmb{Q}_{\text{m}}^{-1} (\frac{\partial h}{\partial \pmb{s}_{k,n}})\right\}+ \pmb{P}^{-1}_{k,n|n-1} .
	\end{align}
	By the definition of PCRB, the MSE matrix of any unbiased estimator satisfies
	\begin{equation} 
		\mathbb{E}
		\!\left[
		(\hat{\pmb{s}}_{k,n}-\pmb{s}_{k,n})
		(\hat{\pmb{s}}_{k,n}-\pmb{s}_{k,n})^{\text H}
		\right]
		\succeq
		\pmb{J}_{k,n}^{-1} = \pmb{C}|_{\pmb{s}_{k,n}},
	\end{equation}
	In practice, since $\pmb{s}_{k,n}$ is typically unknown, $\hat{\pmb{s}}_{k,n|n-1}$ is applied to calculate the predicted FIM, and the corresponding PCRB matrix is denoted as $\pmb{C}|_{\pmb{s}_{k,n}=\hat{\pmb{s}}_{k,n|n-1}}$. 
	Moreover, the PCRB of the $i$-th element belonging to a vector $\pmb{s}_{k,n}$ corresponds to the $i$-th diagonal element of the PCRB matrix of  $\pmb{s}_{k,n}$. Hence, the predicted PCRB of $c^{x}_{k,n}$ and $c^{y}_{k,n}$ can be obtained as
	\begin{align} 
		&\left\{ \begin{array}{l}
		  \text{PCRB}(\hat{c}^{x}_{k,n|n-1})=c_{11}|_{\pmb{s}_{k,n}=\hat{\pmb{s}}_{k,n|n-1}}, \\
		  \text{PCRB}(\hat{c}^{y}_{k,n|n-1})=c_{44}|_{\pmb{s}_{k,n}=\hat{\pmb{s}}_{k,n|n-1}}.
		\end{array}\right.
	\end{align}
	We adopt the sum of $\text{PCRB}(\hat{c}^{x}_{k,n|n-1})$ and $\text{PCRB}(\hat{c}^{y}_{k,n|n-1})$ as the sensing performance metric with respect to buoy $k$ at the $n$-th radio frame, which is given by
	\begin{align}\label{eqTheta} 
		\Theta_{k,n}=(c_{11}+c_{44})|_{\pmb{s}_{k,n}=\hat{\pmb{s}}_{k,n|n-1}} .
	\end{align}
	
	\vspace*{-2mm}
	\subsection{Communication-Related Evaluation}\label{S3.2}
	
	In order to evaluate the performance of communication, the uplink rate (bits/s/Hz) is applied. Based on \eqref{eqtirknt}, the signal-to-noise ratio (SNR) of buoy $k$ at the $(n+1)$-th radio frame can be derived as 
	\begin{align} 
		\text{SNR}_{k,n+1}=
		\frac{P_{\text{b}}\big|\tilde{\alpha}\cdot\varpi_{k,n+1}^{\text{rx}}\big|^2}{{d_{k,n+1}^2 N_0}},
	\end{align}
	where  $d_{k,n+1}=\|\pmb{c}_{\text{U},n}-\pmb{c}_{k,n+1}\|$, and$\varpi_{k,n+1}^{\text{rx}}=\pmb{a}^{\text{H}}_{\text{rx}}(\hat{\pmb{c}}_{k,n+1|n},\pmb{c}_{\text{U},n})\pmb{a}_{\text{rx}}(\pmb{c}_{k,n+1},\pmb{c}_{\text{U},n})$ Then, the uplink rate of buoy $k$ at the $(n+1)$-th radio frame is given by
	\begin{align}\label{eqRate} 
	  R_{k,n+1} = \log_2(1+\text{SNR}_{k,n+1}) .
	\end{align}
	
	\vspace*{-2mm}
	\subsection{Optimization Formulation}\label{S3.3}
	
	According to Fig.~\ref{figure-UUSW-FS}, within each OB, the odd-indexed radio frame $n \in \{2i-1 \mid i=1,2,\ldots,\frac{N}{2}\}$ is used for sensing, state refinement, and proactive beam prediction, whereas the actual buoy uplink transmission takes place in the subsequent $(n+1)$-th radio frame. Accordingly, the sensing objective is evaluated at frame $n$, while the communication-rate requirement is imposed on frame $(n+1)$. Recall that the UAV is assumed to remain stationary within one OB. Therefore, the sensing and communication performances are intrinsically coupled through the common UAV position. Under this coupling, the optimization problem is formulated as:
	\begin{subequations}\label{eq:MDDSCC:Opt} 
		\begin{align}
			(\text{P}1):\quad &\min_{\left\{p_{k,n}\right\}, \left\{c_{\text{U},n}^{x(y)}\right\}}  \max \left\{ \{\Theta_{k,n}\}_{k=1}^K\right\} ,\label{eqOPob}  \\
			\text{s.t.}\quad & R_{k,n+1} \geq R_{\text{s}}, \ \forall k, \label{eqOPob1}\\
			& \sum_{k=1}^K p_{k,n} \leq P_{\text{U}}, \quad p_{k,n} > 0, \ \forall k, \label{eqOPob2} \\
			&\left\|\pmb{c}_{\text{U},n}-\pmb{c}_{\text{U},n-1}\right\|\leq V_{\text{max}}\bar{T}. \label{eqOPob3}
		\end{align} 
	\end{subequations}
	Constraint \eqref{eqOPob1} states that the data rate achieved at the uplink frame of an OB must be larger than a rate threshold $R_{\text{s}}$. Constraint \eqref{eqOPob2} specifies that the total sensing power is limited by the UAV power budget $P_{\text{U}}$\footnote{Here, $P_{\text U}$ denotes the UAV's maximum sensing power rather than the full onboard energy budget. A detailed propulsion-energy model is not considered, since the focus of this paper is on predictive beamforming under buoy dynamics rather than UAV energy management. Moreover, in offshore monitoring networks, UAV missions are often launched from a nearby mother ship with controlled range and duration, so that the available onboard energy is generally sufficient, especially when fuel-powered platforms are used.}. Since $n$ indexes the sensing frame, constraint \eqref{eqOPob3} actually indicates an inter-OB mobility constraint by the maximum UAV speed $V_{\text{max}}$.
	
	\vspace*{-1mm}
	\section{Solution to the Optimization Problem}\label{S4}
	
	\subsection{Reformulation of Objective Function}\label{S4.1}
	
	The optimization (P1) is difficult to solve, as the derivation of $\Theta_{k,n}$ highly depends on the inverse of $\pmb{J}_{k,n}$, i.e., $\Theta_{k,n} =(\pmb{J}_{k,n}^{-1})_{11}+(\pmb{J}_{k,n}^{-1})_{44}$. By defining the selection matrix $\pmb{E}_{14}\! =\! [1,0;0,0;0,0;0,1;0,0;0,0]$, $\Theta_{k,n}$ can be rewritten as
	\begin{equation}\label{eq:Theta_trace} 
		\Theta_{k,n}=\text{tr}\!\left(\pmb E^{\text T}_{14}\pmb J_{k,n}^{-1}\pmb E_{14}\right).
	\end{equation}
	Next we introduce an auxiliary symmetric matrix variable $\pmb U_{k,n} \in \mathbb{S}^{2}_{+}$ and enforce
	\begin{equation}\label{eq:U_bound} 
		\pmb U_{k,n}\succeq \pmb E_{14}^{\text T}\pmb J_{k,n}^{-1}\pmb E_{14},
	\end{equation}
	which implies $\text{tr}(\pmb U_{k,n})\ge \Theta_{k,n}$. Moreover, the expression \eqref{eq:U_bound} can be equivalently written as a linear matrix inequality (LMI) via the Schur complement. Since $\pmb J_{k,n}\succ \pmb 0$, we have
	\begin{equation}\label{eq:LMI_schur} 
		\begin{bmatrix}
			\pmb J_{k,n} & \pmb E_{14}\\
			\pmb E^{\text T}_{14} & \pmb U_{k,n}
		\end{bmatrix}\succeq \pmb 0.
	\end{equation}
	The original objective $\Theta_{k,n}$ in (P1) can be upper-bounded by the linear surrogate $\text{tr}(\pmb U_{k,n})$ under the LMI constraint \eqref{eq:LMI_schur}. In particular, minimizing $\text{tr}(\pmb U_{k,n})$ tightens the upper bound in \eqref{eq:U_bound} and yields an equivalent reformulation in the optimum. To handle the min--max objective over all buoys, we introduce slack variable $\vartheta_n$ and rewrite (P1) as
	\begin{subequations}\label{eq:MDDSCC:Opt2} 
		\begin{align}
			(\text{P}2):	&\min_{\left\{p_{k,n}\right\}, \left\{c_{\text{U},n}^{x(y)}\right\}, \{\pmb U_{k,n}\}} \ \vartheta_n , \\
			\quad\text{s.t.},\quad&\text{tr}(\pmb U_{k,n})\le \vartheta_n,\ \forall k, \label{eq:MDDSCC:Opt2-1} \\
			&\eqref{eqOPob1}, \ \eqref{eqOPob2}, \ \eqref{eqOPob3},\ \eqref{eq:LMI_schur}.
		\end{align}
	\end{subequations}
	Now the objective function is convex, but the constraint \eqref{eq:LMI_schur} is non-convex as $\left\{p_{k,n}\right\}, \left\{c_{\text{U},n}^{x(y)}\right\}$ are tightly coupled in $\pmb{J}_{k,n}$.
	
	\vspace*{-2mm}
	\subsection{Convexification of Constraints}\label{S4.2}
	
	Due to the block-diagonal structure of $\pmb Q_{\text m}$, the measurement information is additive. 
	Let $\pmb H_{k,n}^{(\pmb r)}\triangleq \frac{\partial \pmb r_{k,n}}{\partial \pmb s_{k,n}}$,
	$\pmb h_{k,n}^{(f)}\triangleq \frac{\partial f^{\text D}_{k,n}}{\partial \pmb s_{k,n}}$, and
	$\pmb h_{k,n}^{(\tau)}\triangleq \frac{\partial \tau_{k,n}}{\partial \pmb s_{k,n}}$.
	Then \eqref{eq:FIM_realparam} can be equivalently written as
	\begin{align}\label{eq:FIM_additive} 
		\pmb J_{k,n}
		=
		\sum_{i=1}^{3}\omega_{i,k,n}\pmb A_{i,k,n}+\pmb P^{-1}_{k,n|n-1},
	\end{align}
	where $\omega_{i,k,n}=1/\sigma_{i,k,n}^2$, and the three positive semidefinite (PSD) matrices are defined as
	\begin{align} 
		\pmb A_{1,k,n}&\triangleq 2\,\Re\!\left\{(\pmb H_{k,n}^{(\pmb r)})^{\text H}\pmb H_{k,n}^{(\pmb r)}\right\},\\
		\pmb A_{2,k,n}&\triangleq (\pmb h_{k,n}^{(f)})^{\text T}\pmb h_{k,n}^{(f)},\\
		\pmb A_{3,k,n}&\triangleq (\pmb h_{k,n}^{(\tau)})^{\text T}\pmb h_{k,n}^{(\tau)}.
	\end{align}
	The detailed expressions of these PSD matrices are presented in Appendix \ref{App:rkn}.
	Both $\sigma_{i,k,n}^2$ and PSD matrices contain the optimization variables $p_{k,n}$ and $c_{\text{U},n}^{x(y)}$. Hence, we aim to reform $\pmb{J}_{k,n}$, such that the constraint \eqref{eq:LMI_schur} becomes convex.
	
	Equation \eqref{eq:FIM_additive} contains two coupled sources of non-convexity, namely, the matrix-valued terms $\{\pmb A_{i,k,n}\}$ and the scalar weights $\{\omega_{i,k,n}\}$. To preserve tractability, we adopt a split SCA treatment. Specifically, the extremely complicated matrix terms $\{\pmb A_{i,k,n}\}$ are updated in a lagged manner, whereas the scalar weights $\omega_{i,k,n}=1/\sigma_{i,k,n}^2$ are convexified via auxiliary variables and first-order Taylor expansion (FOTE).
	To this end, we introduce auxiliary variables $\nu_{i,k,n}>0$ to upper-bound the noise variances, i.e.,
	\begin{equation}\label{eq:nu_ge_sigma} 
		\nu_{i,k,n}\ge \sigma_{i,k,n}^2,\quad i\in\{1,2,3\},
	\end{equation}
	and hence $\omega_{i,k,n}=1/\sigma_{i,k,n}^2$ is lower-bounded by $1/\nu_{i,k,n}$. 
	Then, we fix $\{\pmb A_{i,k,n}\}$ as the value of the last iteration, and define the lagged FIM as 
	\begin{align} 
		\bar{\pmb J}_{k,n}^{(\ell)}=
		\sum_{i=1}^{3}\frac{1}{\nu_{i,k,n}}\pmb A_{i,k,n}^{(\ell-1)}+\pmb P^{-1}_{k,n|n-1} .
	\end{align}
	Since $f(\nu)=1/\nu$ is convex on $\nu>0$, 
	its FOTE at the previous iterate $\nu_{i,k,n}^{(\ell-1)}$ yields a global affine lower bound
	\begin{equation}\label{eq:taylor_inv} 
		\frac{1}{\nu_{i,k,n}}
		\ge 
		\underline{\omega}_{i,k,n}^{(\ell)}(\nu_{i,k,n})
		\triangleq
		\frac{1}{\nu_{i,k,n}^{(\ell-1)}}-\frac{\nu_{i,k,n}-\nu_{i,k,n}^{(\ell-1)}}{\left(\nu_{i,k,n}^{(\ell-1)}\right)^{2}}.
	\end{equation}
	
	With the aid of \eqref{eq:taylor_inv},  and by updating $\pmb A_{i,k,n}$ in a lagged manner, we can construct a conservative approximation of the FIM as
	\begin{equation}\label{eq:J_tilde} 
		\widetilde{\pmb J}_{k,n}^{(\ell)}
		=
		\sum_{i=1}^{3}\underline{\omega}_{i,k,n}^{(\ell)}(\nu_{i,k,n})\,\pmb A_{i,k,n}^{(\ell-1)}
		+\pmb P^{-1}_{k,n|n-1}.
	\end{equation}
	Note that $\widetilde{\pmb J}_{k,n}^{(\ell)}\preceq \bar{\pmb J}_{k,n}^{(\ell)}$ holds whenever $\pmb A_{i,k,n}^{(\ell-1)}\succeq\pmb 0$ and $\underline{\omega}_{i,k,n}^{(\ell)}(\nu_{i,k,n})\le 1/\nu_{i,k,n}$. Hence, $\widetilde{\pmb J}_{k,n}^{(\ell)}$ serves as a conservative affine lower bound of the lagged FIM used in the $\ell$-th iteration.
	
	We should also mention that, when $i\in\{2,3\}$, the variance upper-bound in \eqref{eq:nu_ge_sigma} still involves the clutter coefficient $\beta_{k,n}^{\rm clu}$. To keep this constraint convex, we also evaluate $\beta_{k,n}^{\rm clu}$ at the previous iterate, which yields
	\begin{align} \label{eq:nu23} 
		\nu_{i,k,n} &\geq
		{\frac{4c_i^2N_0}{M_f^2|\chi_{k,n}|^2}}
		\frac{\hat{d}_{k,n|n-1}^2}{p_{k,n}}
		+{\frac{4c_i^2\rho_{k,n}^2|\beta^{\text{clu},(\ell-1)}_{k,n}|^2}{|\chi_{k,n}|^2}}
		\hat{d}_{k,n|n-1}^2, \nonumber \\
		&i\in\{2,3\},
	\end{align}
	where $\hat{d}_{k,n|n-1}=\|\pmb c_{\text U,n}-\hat{\pmb c}_{k,n|n-1}\|$, with $\hat{\pmb c}_{k,n|n-1}$ treated as known for design. Since the right-hand side of the inequality is the sum of two convex functions, \eqref{eq:nu23} is a convex constraint.
	
	\begin{remark}\label{remark:full} 
		The main purpose of freezing $\{\pmb A_{i,k,n}\}$ in \eqref{eq:J_tilde} is to keep each SCA subproblem simple and computationally efficient. An alternative is to apply Full-FOTE to all the non-convex terms in $\pmb J_{k,n}$, which has been widely-used in ISAC-related designs \cite{Jing2024Sky,Lyu2023Maneuver}. Although such a treatment can reduce the approximation loss, it also requires recomputing the linearization of the entire FIM for all buoys at every iteration, and usually incurs a noticeably higher practical optimization cost. By contrast, the adopted lagged-SCA strategy updates $\{\pmb A_{i,k,n}\}$ and $\beta_{k,n}^{\rm clu}$ only through the previous iterate, while retaining a conservative convex surrogate in the current step.
		This design is well suited to the considered beam-tracking regime. In particular, the EKF prediction keeps the pointing mismatch within the mainlobe over consecutive updates, so that the frozen quantities vary smoothly across iterations. A standard safeguard to further stabilize the lagged approximation is to employ a trust-region update of $\pmb c_{\text U,n}$, i.e., 
		\begin{align}\label{eqcUnl} 
			\|\pmb c_{\text U,n}^{(\ell)}-\pmb c_{\text U,n}^{(\ell-1)}\|\le \delta^{(\ell)},
		\end{align}
		which improves numerical robustness. A performance-complexity comparison between the proposed lagged-SCA method and Full-FOTE will be presented in Section~\ref{S5}.
	\end{remark}

	The dominant non-convexity in the uplink-rate constraint \eqref{eqOPob1} comes from the receive-beam mismatch factor. In the considered short-horizon beam-tracking regime, the resulting rate variation is mainly driven by the prediction accuracy of the buoy position, whereas the direct effect of optimizing $\pmb c_{\text U,n}$ on the mismatch factor is relatively mild. Moreover, the refined predictor $\hat{\pmb c}_{k,n+1|n}$ becomes available only after processing the $n$-th sensing echoes, while Problem (P2) must be solved beforehand. To preserve causality, we retain the path-loss dependence through the available two-step predictor $\hat{\pmb c}_{k,n+1|n-1}$ and assume the receive-mismatch factor $\varpi_{k,n+1}^{\text{rx}}\! \approx\! 1$.\footnote{The realized uplink-rate constraint may be occasionally violated due to the omitted mismatch factor and prediction uncertainty, especially under harsher sea conditions, which will be analyzed in Fig.~\ref{figure-UUSW-ss}.} Accordingly, constraint \eqref{eqOPob1} is approximated by
	\begin{align} \label{eqdkn22} 
		\hat d_{k,n+1|n-1}^2 \leq \frac{P_{\text b}|\tilde{\alpha}|^2}{N_0 (2^{R_{\text{s}}}-1 )},
	\end{align}
	with $\hat d_{k,n+1|n-1}\triangleq\|\pmb c_{\text U,n}-\hat{\pmb c}_{k,n+1|n-1}\|$, which is a standard second-order cone (SOC) constraint in $\pmb c_{\text U,n}$.
	Then, at the $\ell$-th iteration, Problem (P2) is handled by solving the following convex problem
	\begin{subequations}\label{eq:P2_SDP} 
		\begin{align}
			&(\text{P3}):\quad \min_{\{p_{k,n}\},\,\pmb c_{\text U,n},\,\{\pmb U_{k,n}\},\,\{\nu_{i,k,n}\}}\quad  \vartheta_n  ,\label{eq:P2_SDP_obj} \\
			\text{s.t.}\quad &\eqref{eqOPob2}, \eqref{eqOPob3}, \eqref{eq:MDDSCC:Opt2-1}, \eqref{eq:nu23}, \eqref{eqcUnl}, \eqref{eqdkn22} \label{eq:P2_SDP_rate}, \\
			& \begin{bmatrix}
				\widetilde{\pmb J}_{k,n}^{(\ell)} & \pmb E_{14}\\
				\pmb E_{14}^{\text T} & \pmb U_{k,n}
			\end{bmatrix}\succeq \pmb 0,\ \forall k, \label{eq:P2_SDP_LMI} \\
			&\underline{\omega}_{i,k,n}^{(\ell)}(\nu_{i,k,n}) \geq 0, \forall i, \ k, \label{eq52d} \\
			& \nu_{1,k,n} \geq {c_1^2 N_0}/({{p_{k,n}}M_{\text{f}}^2}),\ \forall k. \label{eq52e}
		\end{align}
	\end{subequations}
	
	\begin{algorithm}[!t]
		\caption{Lagged-SCA Based Solution to (P1) at Radio Frame $n$}\label{alg:SCA_P2}
		\KwIn{Predicted buoy states $\{\hat{\pmb c}_{k,n|n-1}\}$, prior covariance $\{\pmb P_{k,n|n-1}\}$, previous UAV position $\pmb c_{\text U,n-1}$, budgets $P_{\text U}$ and $V_{\max}$, rate threshold $R_{\text s}$, tolerance $\epsilon$, maximum iterations $I_{\max}$, trust-region radius $\delta$.}
		\KwOut{Power allocation $\{p_{k,n}^\star\}$ and UAV position $\pmb c_{\text U,n}^\star$.}
		
		\BlankLine
		\textbf{Initialization:} Set $\ell=0$. Choose a feasible $\{p_{k,n}^{(0)}\}$ and $\pmb c_{\text U,n}^{(0)}$. 
		Compute $\beta_{k,n}^{\text{clu},(0)}$ and $\pmb A_{i,k,n}^{(0)}$ based on $\pmb c_{\text U,n}^{(0)}$ and $\hat{\pmb c}_{k,n|n-1}$. 
		Set feasible $\{\nu_{i,k,n}^{(0)}\}$ and $\vartheta_n^{(0)}$\;
		
		\For{$\ell=1$ \KwTo $I_{\max}$}{
			Update frozen quantities $\{\pmb A_{i,k,n}^{(\ell-1)}\}$ and $\{\beta_{k,n}^{\text{clu},(\ell-1)}\}$ using $\pmb c_{\text U,n}^{(\ell-1)}$ and $\hat{\pmb c}_{k,n|n-1}$\;
			Construct affine lower bounds $\underline{\omega}_{i,k,n}^{(\ell)}(\nu_{i,k,n})$ via \eqref{eq:taylor_inv}\;
			Form $\widetilde{\pmb J}_{k,n}^{(\ell)}$ via \eqref{eq:J_tilde}\;
			Solve the optimization problem (P3) to obtain
			$\{p_{k,n}^{(\ell)},\pmb c_{\text U,n}^{(\ell)},\pmb U_{k,n}^{(\ell)},\nu_{i,k,n}^{(\ell)},\vartheta_n^{(\ell)}\}$\;
			\If{$\left|\vartheta_n^{(\ell)}-\vartheta_n^{(\ell-1)}\right|/\max\{1,\vartheta_n^{(\ell-1)}\}\le \epsilon$}{
				\textbf{break}\;
			}
		}
		Set $\{p_{k,n}^\star\}=\{p_{k,n}^{(\ell)}\}$ and $\pmb c_{\text U,n}^\star=\pmb c_{\text U,n}^{(\ell)}$.
	\end{algorithm}
	
	\vspace*{-2mm}
	\subsection{Algorithm and Its Computational Complexity}\label{S4.3}

		The overall optimization procedure is summarized in Algorithm~\ref{alg:SCA_P2}, the computational complexity of which is discussed as follows. For fixed state dimension and antenna size, Steps 3--5 of Algorithm~\ref{alg:SCA_P2} only update the lagged quantities and affine lower bounds, and thus incur $\mathcal{O}(K)$ arithmetic cost. The dominant complexity comes from solving the optimization (P3) in Step 6. Specifically, (P3) contains $K$ LMIs of size $8\times 8$, and the number of scalar decision variables is $n_{\mathrm v} = K\! +\! 3K\! +\! 3K\! +\! 2\! +\! 1 = \mathcal{O}(K)$, where the terms correspond to $\{p_{k,n}\}$, $\{\nu_{i,k,n}\}$, $\{\pmb U_{k,n}\}$, $\pmb c_{\text U,n}$, and $\vartheta_n$, respectively.
		According to the standard complexity analysis of primal--dual interior-point methods for semidefinite programs \cite{boyd2004convex,vandenberghe1996semidefinite}, the per-step arithmetic cost is upper bounded by $\mathcal{O}\!\left(n_{\mathrm v}^2 n_{\mathrm s}+n_{\mathrm v} n_{\mathrm s}^2+n_{\mathrm s}^3\right)$, where $n_{\mathrm s}=8K$ is the total semidefinite block dimension. Hence, the per-step cost scales as $\mathcal{O}(K^3)$. Let $I_{\mathrm{IP}}$ and $I_{\mathrm{SCA}}$ denote the numbers of interior-point and outer SCA iterations, respectively. Then the overall worst-case complexity of Algorithm~\ref{alg:SCA_P2} is $\mathcal{O}\!\left(I_{\mathrm{SCA}}I_{\mathrm{IP}}K^3\right)$.
		The remaining constraints, namely, the linear constraints (\ref{eqOPob2}), (\ref{eq:MDDSCC:Opt2-1}) and (\ref{eq52d}), the SOC constraints (\ref{eqOPob3}), (\ref{eqdkn22}) and (\ref{eqcUnl}), as well as the rotated-SOC constraints in (\ref{eq:nu23}) and (\ref{eq52e}), only contribute lower-order complexity and hence do not change the dominant cubic scaling with respect to $K$.
	
	\vspace*{-1mm}
	\section{Simulation Results}\label{S5}
	
	This section presents simulation results for the proposed UAV-enabled ocean monitoring network under practical sea effects.
	The mission period consists of $N=100$ radio frames grouped into $N_{\text{block}}=50$ OBs, and each radio frame lasts 0.5\,s. 
	The monitored sea area is centered at $[0,0]^{\text T}$ and spans $400\,\text{m}\times 400\,\text{m}$. We consider $K\! =\! 9$ surface buoys randomly distributed in the sea area, while the UAV starts from $[120,120,50]^{\text T}\,\text m$ and flies at a fixed altitude of $50\,\text m$ with maximum speed $V_{\max}\! =\! 30\,\text{m/s}$. The UAV is equipped with an $8\times 8$ UPA and operates at carrier frequency $f_{\text c}\! =\! 18\,\text{GHz}$ with bandwidth $B\! =\!100\,\text{MHz}$. Unless otherwise stated, the remaining default parameters are summarized in Table~\ref{tab:sim_param_default}. In order to clearly demonstrate the system performance, we adopt the predicted horizontal position root MSE (RMSE) and uplink rate of the worst-buoy as sensing and communication metrics, respectively\footnote{The predicted PCRB-based metric $\Theta_{k,n}$ is used for optimization, whereas the worst-buoy predicted RMSE is applied to evaluate the realized beam-prediction accuracy in simulation.}. In particular, in order to be consistent with uplink rate at the $n+1$ frame, the result of OSP from the sensing frame to communication frame (i.e., $\hat{\pmb{c}}_{n+1|n}$) is used to compute the RMSE. 
	
	\begin{table}[t]
		\caption{Default simulation parameters}
		\label{tab:sim_param_default} 
		\vspace*{-1mm}
		\centering
		\footnotesize
		\renewcommand{\arraystretch}{1.08}
		\setlength{\tabcolsep}{6pt}
		\begin{tabular}{|c|c|c|c|}
			\hline
			\textbf{Parameter} & \textbf{Value} & \textbf{Parameter} & \textbf{Value} \\ \hline
			$P_{\text U}$ & $36~\text{dBm}$ & $P_{\text b}$ & $20~\text{dBm}$ \\ \hline
			$N_0$ & $-174~\text{dBm/Hz}$ & $R_{\text s}$ & $5~\text{bit/s/Hz}$ \\ \hline
			$\kappa_{\text s}$ & $5$ & $\rho$ & $0.2$ \\ \hline
			$\varsigma$ & $1$ & $\sigma_a$ & $0.3~\text{m/s}^2$ \\ \hline
			$\pmb v_{\text c}$ & $[0.4,0.2]^{\text T}~\text{m/s}$ & $(c_1,c_2,c_3)$ & $(1,0.1,0.1)$ \\ \hline
		\end{tabular}
	\end{table}

	\begin{figure}[!h]
		\vspace*{-1mm}
		\centering
		\includegraphics[width=0.95\linewidth]{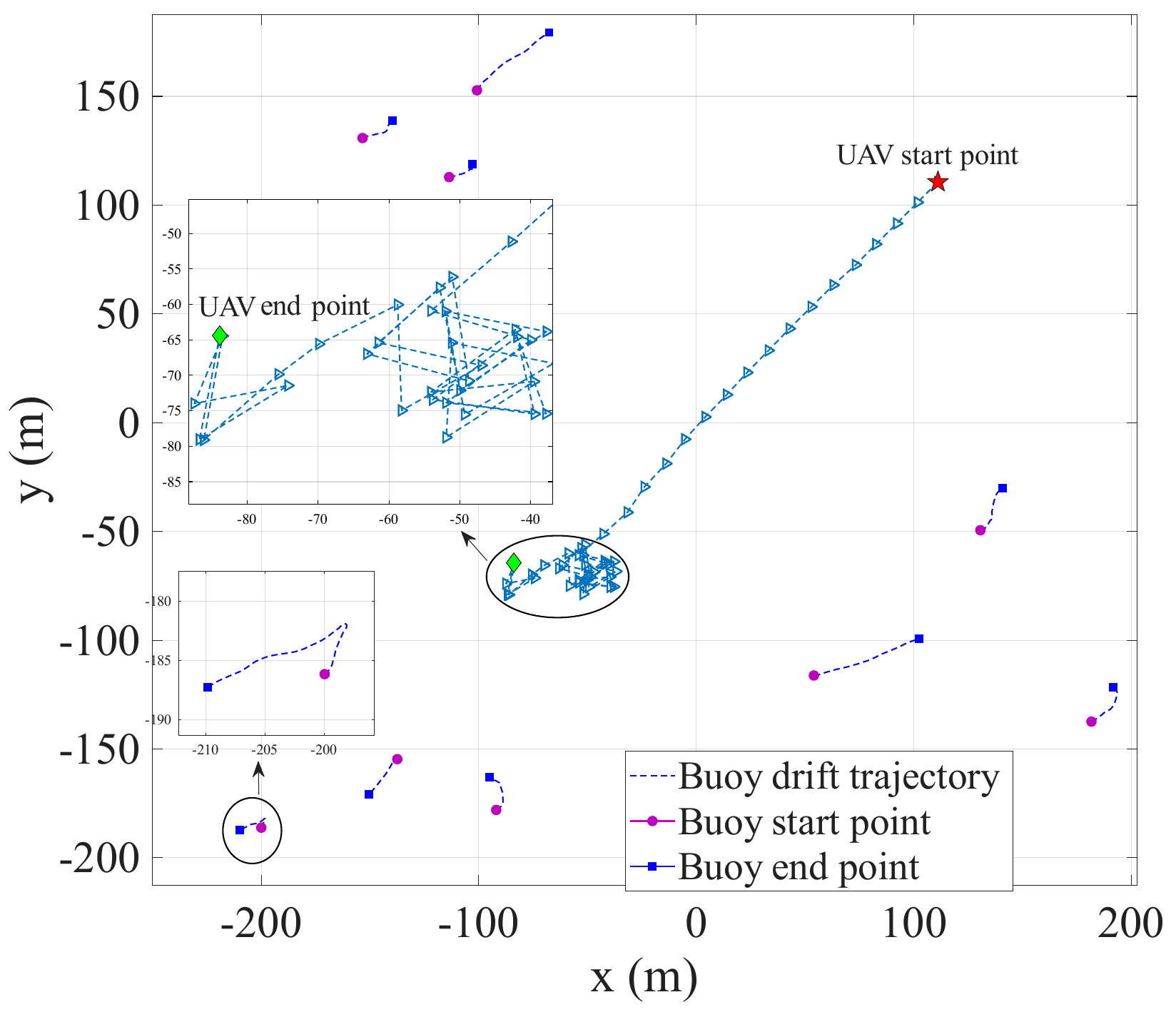}
		\vspace*{-2mm}
		\caption{\small 2D trajectories of the UAV and buoys under one network realization.}
		\label{figure-UUSW-2DT} 
		\vspace*{-1mm}
	\end{figure}  
	
		\vspace*{-2mm}
	\subsection{Performance Results in Proposed Scenario}\label{S5.1}
	
	Fig. \ref{figure-UUSW-2DT} depicts the 2D trajectories of the buoys and UAV over one 100-frame mission. It can be observed that although the ocean current drives the buoys toward the positive directions of the x- and y-axes on average, wave excitation still causes irregular local deviations, and some buoys may even exhibit temporary counter-current motion. 
	The UAV first performs a relatively long transit toward the remote part of the monitored region and then loiters within a compact area.
	This behavior indicates that the optimized trajectory is not intended to follow any individual buoy. Instead, the UAV gradually moves toward a favorable compromise location for the entire buoy set, which is consistent with the proposed worst-case design objective under the mobility, power, and rate constraints.
	
	\begin{figure}[!t]
		\vspace*{-3mm}
		\centering
		\includegraphics[width=0.95\linewidth]{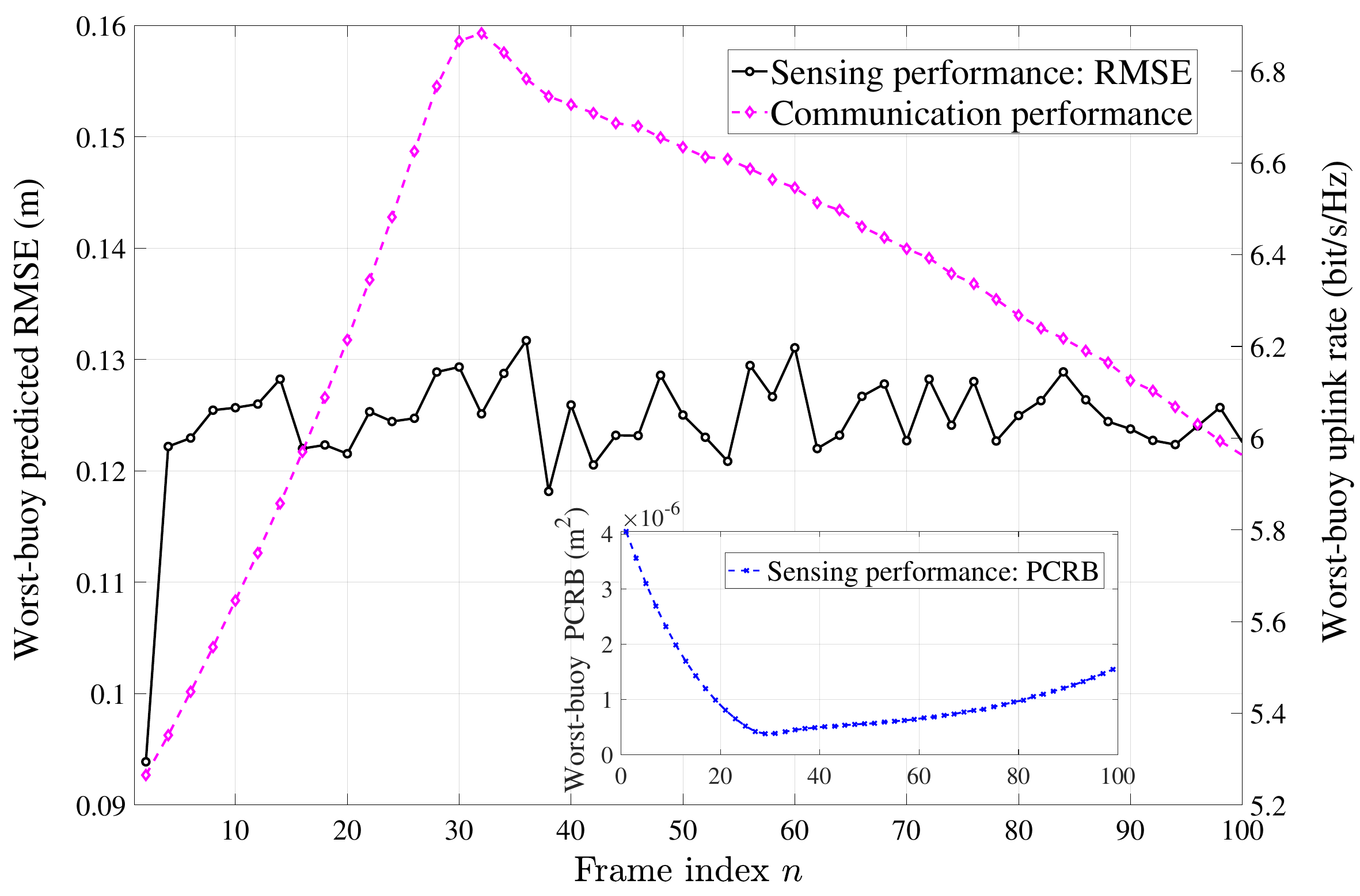}
		\vspace*{-2mm}
		\caption{\small Variation of sensing and communication performance within one 100-frame mission period.}
		\label{figure-UUSW-metric} 
	\end{figure}  
		
	\begin{figure}[!b]
		\vspace*{-1mm}
		\begin{center}
		\includegraphics[width=0.99\linewidth]{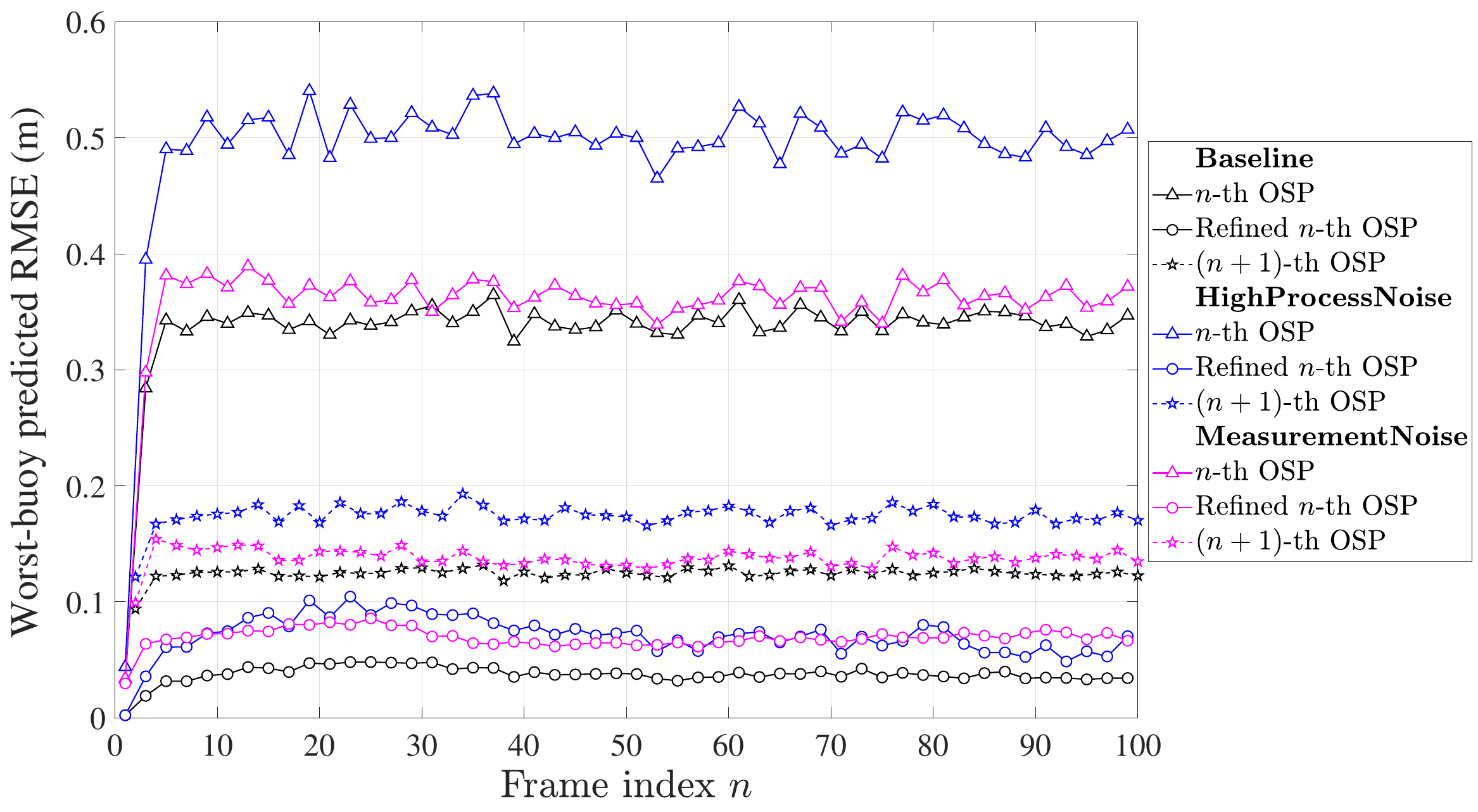}
		\end{center}
		\vspace*{-4mm}
		\caption{\small Sensing performance under different process- and measurement-noise settings.}
		\label{figure-UUSW-rmse} 
	\end{figure} 
		
	Following the same initial states and network layout,  the Monte Carlo simulation using 100 channel realizations is applied to analyze the frame-level performance in Figs. \ref{figure-UUSW-metric} and \ref{figure-UUSW-rmse}. 
	As shown in Fig.~\ref{figure-UUSW-metric}, the sensing-side RMSE and PCRB jointly reveal a transition from an initialization-dominated transient to a steady tracking regime. At the beginning of the mission, the predicted RMSE is exceptionally small because the estimator starts from a nearly perfect initialization, whereas the PCRB is relatively loose since the initial UAV position is not yet sensing-optimal for the worst buoy. As the UAV moves toward a more favorable compromise region, the sensing geometry improves and the PCRB decreases rapidly. Meanwhile, the benefit of the ideal initialization gradually fades, and the accumulated wave-driven motion uncertainty pushes the realized RMSE upward. After this transient stage, both metrics enter a stable operating regime: the RMSE fluctuates around a nearly constant level, while the PCRB increases mildly after reaching its minimum. This indicates that the initial geometric gain has gradually saturated, and the continuing buoy drift together with prediction uncertainty slightly weakens the worst-buoy sensing condition in the later stage.
	In contrast, the communication rate first increases because the UAV moves from its initial location toward a more favorable region for the buoy set, which shortens the worst-buoy uplink distance and improves the overall geometry. After the UAV reaches this compromise region, the geometric gain becomes limited, while the continuing buoy drift and prediction mismatch slightly weaken the worst-buoy beam alignment, leading to a mild rate decrease in the later stage.

	To further illustrate the robustness of the proposed prediction framework under higher measurement noise and stronger Singer process noise, we introduce two additional comparison cases besides the baseline setting. In the first case, termed `HighProcessNoise', the Singer-model process-noise parameter $\sigma_a$ is increased from 0.3 to 0.45 to emulate more violent buoy motion induced by stronger environmental dynamics. In the second case, termed `MeasurementNoise', the measurement-related parameters $c_i$ are set to $(50,5,5)$ to represent degraded sensing quality. 
	As shown in Fig. \ref{figure-UUSW-rmse}, the baseline case achieves the best prediction accuracy at all prediction stages. When the process noise is enlarged, the prediction performance degrades noticeably, indicating that stronger motion uncertainty directly weakens short-term state predictability. When the measurement noise is increased, the degradation is relatively milder, but the benefit of EKF refinement is still reduced because the sensing update becomes less informative. Nevertheless, in all three cases, the refined OSP (i.e., $\hat{\pmb{c}}_{n}$) remains consistently more accurate than the OSP (i.e., $\hat{\pmb{c}}_{n|n-1}$ and $\hat{\pmb{c}}_{n+1|n}$), which verifies the effectiveness of sensing-assisted state refinement under both dynamic and measurement degradations.  
	
	\begin{figure}[!h]
		\vspace*{-1mm}
		\begin{center}
		\includegraphics[width=0.95\linewidth]{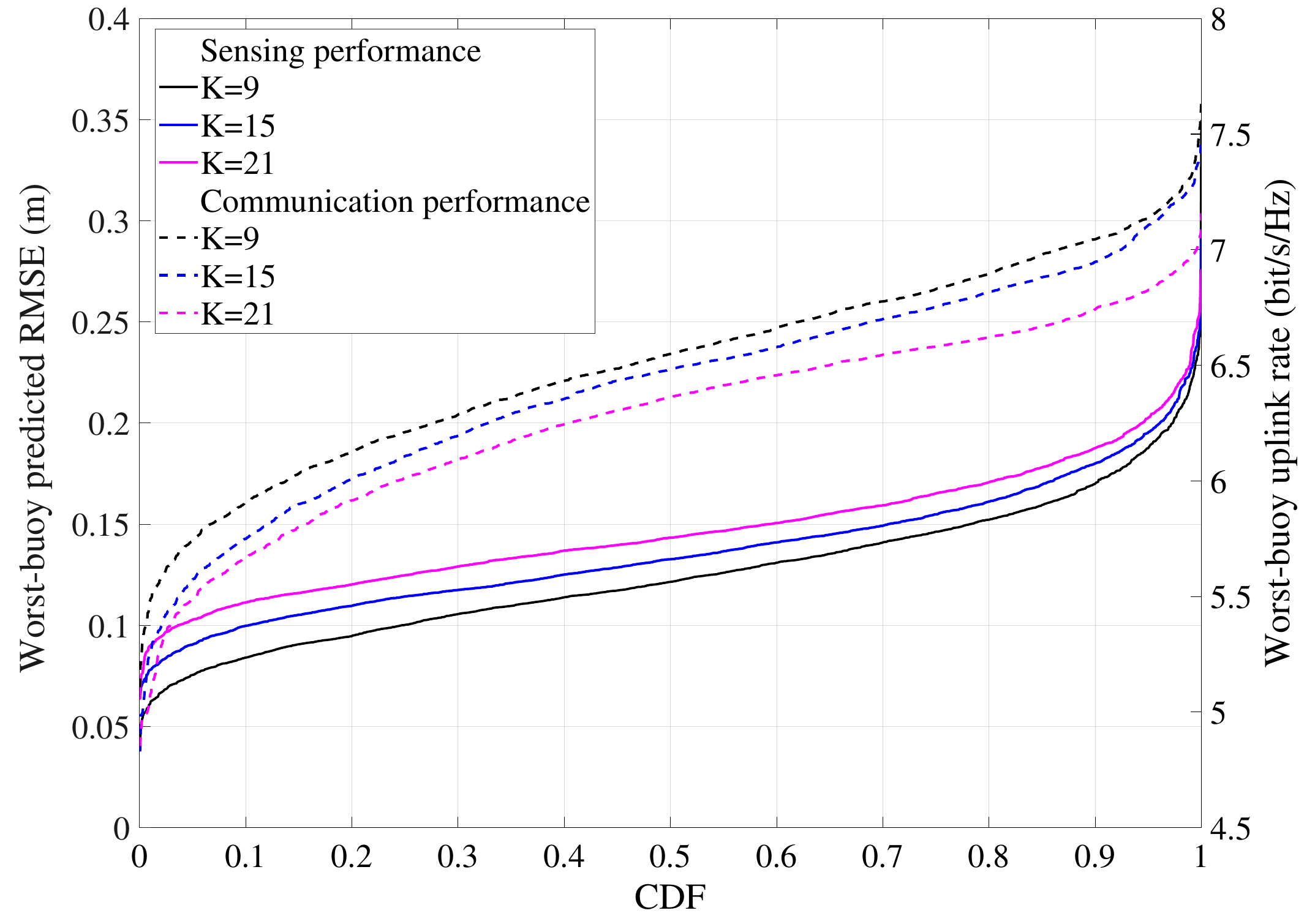}
		\end{center}
		\vspace*{-5mm}
		\caption{\small Sensing and communication performance under different numbers of buoys.}
		\label{figure-UUSW-kcdf} 
		\vspace*{1mm}
	\end{figure}  
	
	To evaluate the system performance under different buoy distributions, the empirical cumulative distribution function (CDF) is plotted using 100 random network realizations.
	Fig. \ref{figure-UUSW-kcdf} compares the worst-buoy sensing and communication performance under different buoy numbers. As the number of buoys increases, the sensing curves move toward less favorable values, while the communication curves move toward lower rates, indicating that the worst buoy becomes harder to support as the network becomes denser. This is because within the same monitored area, the UAV has to balance a larger and more spatially dispersed buoy set under the same mobility, power and rate constraints, and the geometry experienced by the worst buoy becomes less favorable. Nevertheless, the overall trends remain stable, which suggests that the proposed method preserves its effectiveness even though the worst-case operating condition becomes more demanding.
	
	\begin{figure}[!t]
		\vspace*{-1mm}
		\begin{center}
		\includegraphics[width=0.95\linewidth]{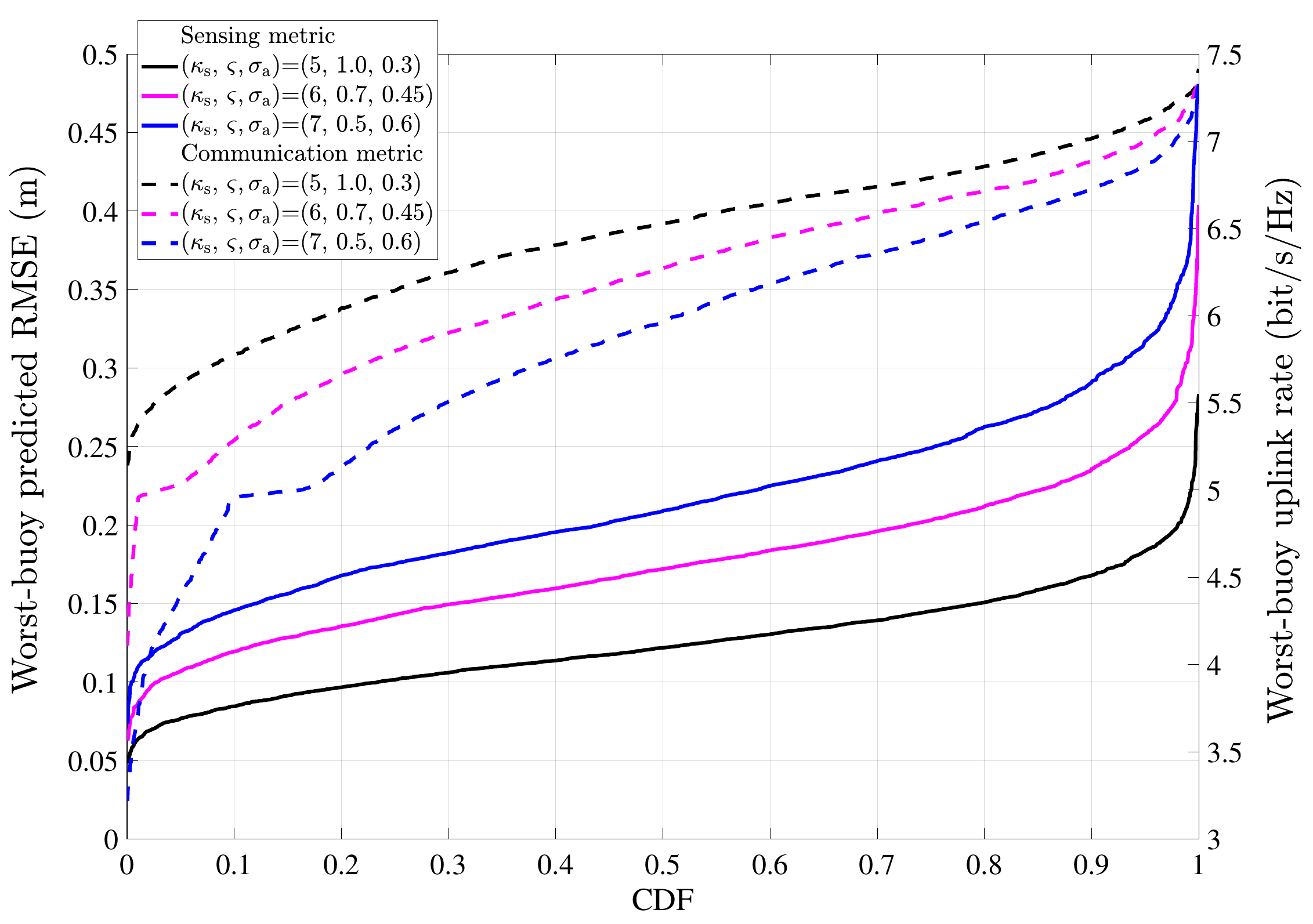}
		\end{center}
		\vspace*{-5mm}
		\caption{\small Sensing and communication performance under different sea conditions.}
		\label{figure-UUSW-ss} 
	\end{figure}  
	
	To emulate different sea states in a physically consistent manner, the sea-state index $\kappa_s$ in the clutter model is jointly varied with the Singer-model parameters $(\varsigma,\sigma_a)$ that characterize the buoy motion.
	The parameter selection follows the underlying physical trends of ocean waves. As the sea state becomes severer, the buoy is subject to stronger wave-induced excitation, while the dominant wave period typically increases, leading to a longer temporal correlation in the motion dynamics \cite{fossen2021handbook}. Accordingly, the acceleration correlation rate $\varsigma$ is decreased, whereas the acceleration standard deviation $\sigma_a$ is increased with $\kappa_s$. In particular, $\sigma_a$ is determined based on the empirical scaling of $\sigma_a = c_a \omega_p^2 H_s$ \cite{holthuijsen2007waves}, where $H_s$ denotes a representative significant wave height, $\omega_p$ is the dominant wave angular frequency, and $c_a=0.12$ is an empirical coefficient accounting for the mapping from wave intensity to horizontal buoy excitation. Following this principle, the parameter triples $(\kappa_s,\varsigma,\sigma_a)$ are set to $(5,1.0,0.30)$, $(6,0.7,0.45)$ and $(7,0.5,0.6)$, such that both the sea clutter level and the wave-induced motion uncertainty increase consistently with the sea severity. Fig.~\ref{figure-UUSW-ss} investigates the impact of sea conditions on the proposed system. 
	
	It can be observed from Fig.~\ref{figure-UUSW-ss} that, as the sea condition becomes more severe, the RMSE curves shift toward larger errors, while the communication curves shift toward lower rates. This indicates that harsher wave-induced dynamics make the buoy motion harder to predict and weaken the effectiveness of sensing-assisted state refinement. As a result, the predicted buoy states used for the subsequent uplink frame become less accurate, which in turn leads to poorer receive-beam alignment at the UAV and lower uplink reliability. It is also worth noting that under the harsher sea-state settings, the lower tail of the worst-buoy uplink-rate CDF falls below the target threshold $R_{\text{s}}=5$ bit/s/Hz. This is because as the sea condition becomes more severe, the combined effects of stronger wave-induced motion uncertainty and higher residual sea clutter enlarge the beam-prediction mismatch, so that the realized uplink rate may occasionally drop below the  threshold. Nevertheless, the performance degradation is gradual across different sea states, rather than abrupt, which suggests that the proposed method degrades gracefully as the maritime environment becomes harsher.
	
	\begin{figure}[!t]
		\vspace*{-1mm}
		\begin{center}
		\includegraphics[width=0.95\linewidth]{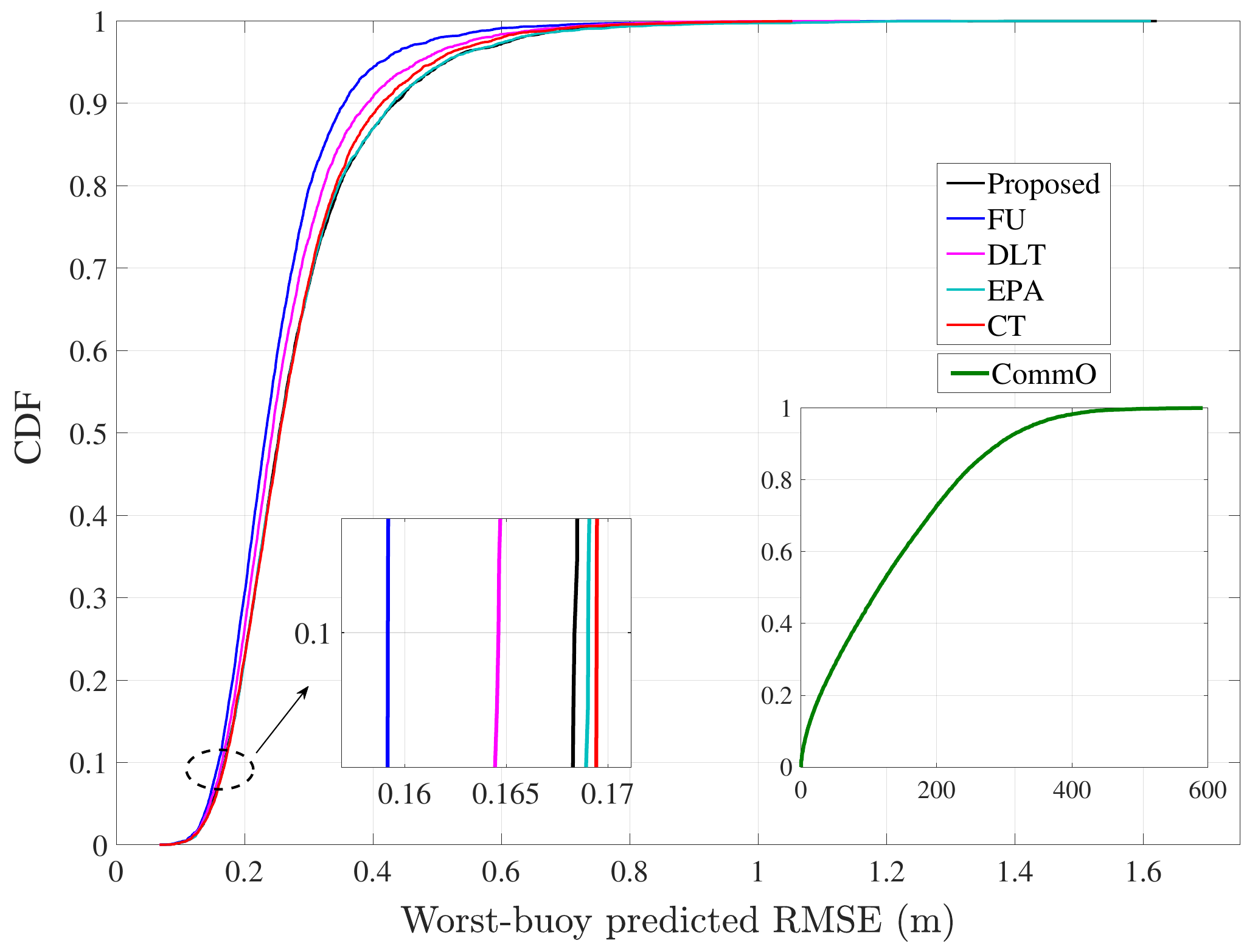}
		\end{center}
		\vspace*{-4mm}
		\caption{\small Sensing performance comparison between the proposed scheme and five benchmarks in terms of the worst-buoy predicted RMSE.}
		\label{figure-UUSW-brmse} 
		\vspace*{0mm}
	\end{figure}  
	
	\begin{figure}[!b]
		\vspace*{-1mm}
		\begin{center}
		\includegraphics[width=0.95\linewidth]{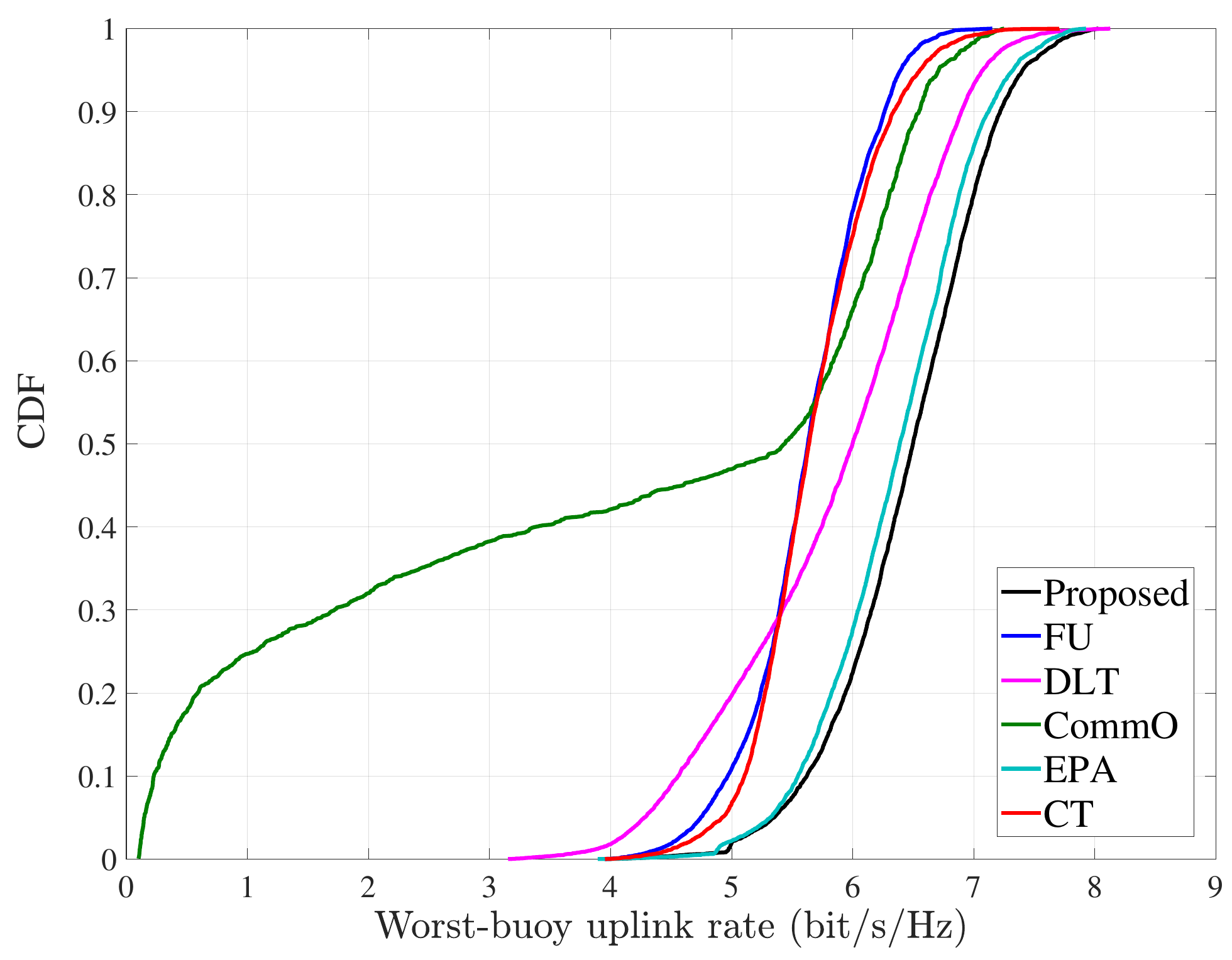}
		\end{center}
		\vspace*{-4mm}
		\caption{\small Communication performance comparison between the proposed scheme and five benchmarks in terms of the worst-buoy uplink rate.}
		\label{figure-UUSW-brate} 
		\vspace*{-1mm}
	\end{figure} 
	
	\vspace*{-2mm}
	\subsection{Performance Comparison with Benchmarks}\label{S5.2}
	
	To highlight the advantages of the proposed sensing-assisted beam prediction approach in the UAV--buoy maritime communications, we compare the proposed design with the following benchmarks under a more challenging setting, namely, $K=15, \bar{R}=16 \times 16$ and $(\kappa_s,\varsigma,\sigma_a) = (6, 0.7, 0.45)$.
	\begin{itemize}
		\item{Equal Power Allocation (\bf {EPA})}: The UAV trajectory is optimized as in the proposed method, but the sensing power is forced to be uniformly allocated among all buoys at every sensing frame, namely, $p_{k,n}=P_{\text U}/K, \ \forall n$.
		\item{Fixed UAV (\bf {FU})}: The UAV remains at its initial position throughout the entire mission period.
		\item{Communication-Only (\bf{CommO})}: The sensing update is disabled, and beam prediction relies only on open-loop state propagation. 
	\end{itemize}
	In addition, inspired by \cite{Jing2024Sky}, we present two low-complexity trajectory schemes, in which a straight- or circular-based UAV trajectory is predefined under the same velocity constraint and only the sensing power is optimizable when solving (P1) problem during mission period. 
	\begin{itemize}
		\item{Diagonal-Line Trajectory (\bf{DLT})}: The UAV flies along the line connecting the initial point and the point of $[-200,-200]^{\text T}$. 
		\item{Circular Trajectory (\bf{CT})}: The UAV follows a preset circular waypoint sequence. The center of the circle is vertically below that initial point, and its radius  is chosen such that the distance between two adjacent waypoints satisfies the mobility constraint. 
	\end{itemize}
	
	The sensing and communication performance comparisons between the proposed scheme and five benchmarks are presented in Figs. \ref{figure-UUSW-brmse} and \ref{figure-UUSW-brate}, respectively. It can be observed that the proposed method provides the most balanced overall performance, since it combines sensing refinement with online UAV trajectory adaptation and is therefore better able to maintain both prediction quality and communication geometry for the worst buoy. In contrast, the FU and CommO suffer clear degradation, because the former loses mobility adaptation while the latter accumulates beam-prediction mismatch without sensing-assisted correction. The predefined-trajectory benchmarks, i.e., DLT and CT, remain more stable than the FU and CommO baselines, but their fixed motion patterns cannot continuously track the time-varying buoy distribution, and hence they are still inferior to the proposed design. In addition, the EPA remains relatively competitive, which suggests that trajectory adaptation is the main factor in preserving the worst-buoy performance, while adaptive sensing-power allocation mainly provides an additional robustness gain.
	
	\begin{figure}[!t]
		\vspace*{-1mm}
		\begin{center}
		\includegraphics[width=0.99\linewidth]{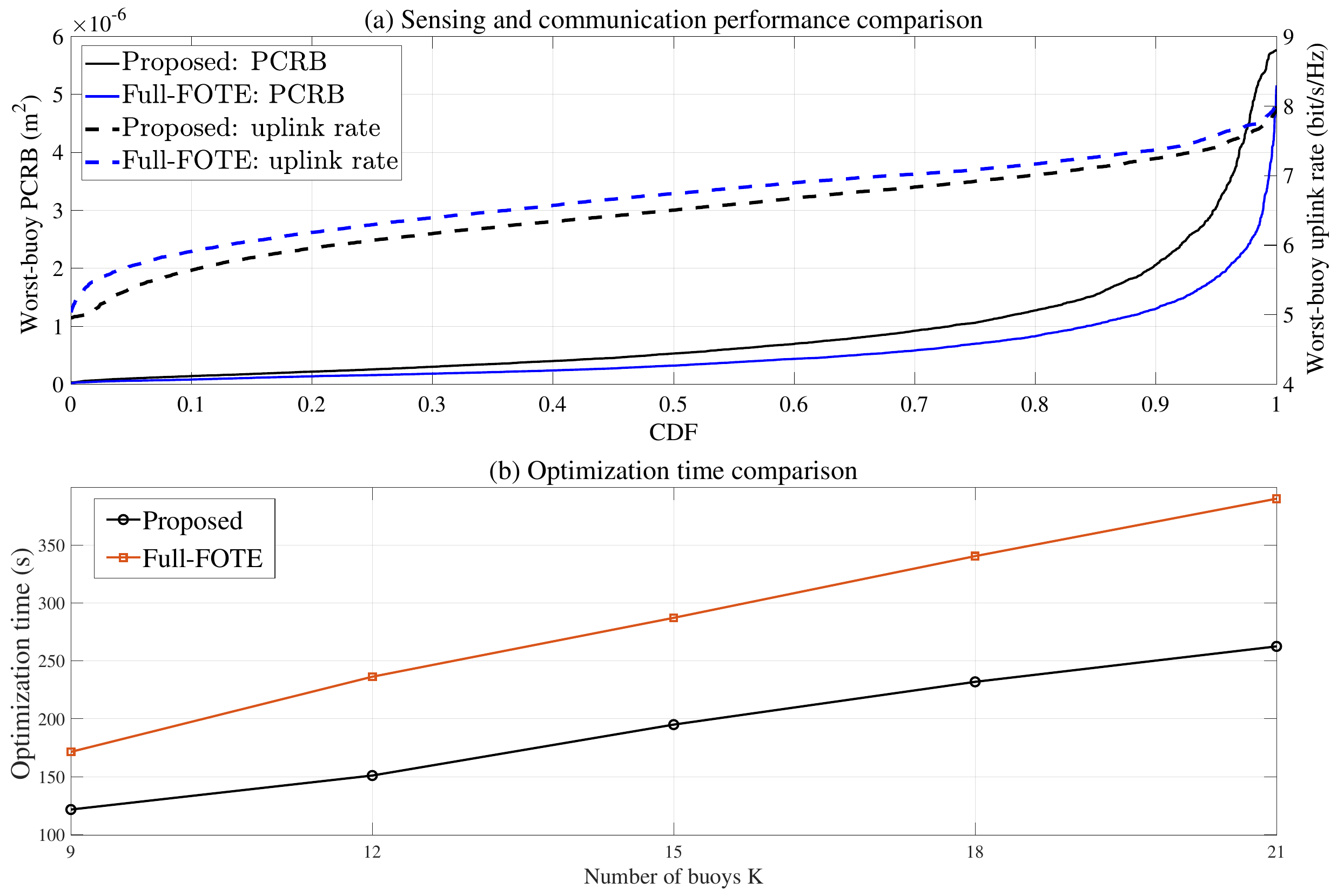}
		\end{center}
		\vspace*{-4mm}
		\caption{\small Performance and optimization-time comparison between the proposed lagged-SCA scheme and the Full-FOTE benchmark. The optimization time, averaged over 100 random realizations, is measured on a desktop with AMD Ryzen 9 9950X3D 16-core CPU, MATLAB R2024b, CVX 2.2.}
		\label{figure-UUSW-fullT} 
		\vspace*{1mm}
	\end{figure} 
	
	Recall that to reduce the computational burden of solving (P1), the proposed method adopts a lagged-SCA treatment, where the matrices $\pmb A_{i,k,n}$ in \eqref{eq:FIM_additive} are updated using the previous iterate, while the scalar weights $\omega_{i,k,n}$ are handled via FOTE, leading to the affine FIM surrogate in \eqref{eq:J_tilde}. Although this treatment greatly simplifies each SCA subproblem, it may introduce some approximation loss compared with the Full-FOTE mentioned in Remark \ref{remark:full}. In the following, we compare the proposed method with Full-FOTE in terms of both system performance and optimization time.
	
	As shown in Fig.~\ref{figure-UUSW-fullT}, Full-FOTE indeed achieves a lower sensing cost function in (P1) and also yields slightly higher communication rates, which confirms that a more accurate first-order treatment of the entire FIM can further reduce the approximation loss introduced by the proposed lagged-SCA method. Nevertheless, this performance improvement is moderate compared with the additional optimization complexity. In particular, as shown in Fig.~\ref{figure-UUSW-fullT}\,(a), the 90\%-likely uplink rate achieved by Full-FOTE is only 0.27 bit/s/Hz higher than that of the proposed method. By contrast, Fig.~\ref{figure-UUSW-fullT}\,(b) shows that the optimization-time gap becomes more pronounced as the number of buoys $K$ increases. Especially when $K=21$, Full-FOTE requires nearly 150\,s more to complete a 100-frame optimization mission. This is because the proposed method only updates the frozen matrix terms and solves the convex subproblem (P3) at each iteration of Algorithm~1. By contrast, Full-FOTE needs to recompute the first-order linearization of the entire FIM for all buoys at every iteration, and typically requires additional trust-region safeguards to maintain numerical stability, which further increases the number of iterations. Therefore, although Full-FOTE serves as a useful upper-performance reference, the proposed method offers a more attractive balance between performance and computational complexity for online deployment.
	
	\vspace*{-2mm}
	\section{Conclusions}\label{S6}
	
	This paper has investigated sensing-assisted predictive beamforming for UAV-enabled ocean monitoring networks. By incorporating residual sea clutter, wave-induced motion and current-induced drift into an ISAC-based UAV mission workflow, we have developed an EKF prediction/refinement framework and derived a PCRB-based sensing metric for proactive beam alignment. Based on this metric, we have formulated a worst-buoy joint sensing-power and UAV-mobility optimization problem under communication, power, and mobility constraints, and have solved its per-frame surrogate through an LMI-based lagged-SCA algorithm with convex-conic complexity. Numerical results have demonstrated that the proposed design provides robust mission-long performance and remains effective when the buoy density, process uncertainty, measurement noise, and sea severity increase. It also achieves the most balanced sensing-communication performance among the considered benchmarks. The marked RMSE degradation of the communication-only benchmark further confirms that sensing-assisted state refinement is indispensable for reliable predictive beamforming under sea-surface dynamics. Moreover, although the Full-FOTE benchmark yields further performance gains, its computational burden is substantially higher, which highlights the advantage of the proposed lagged design.
	
	Future work will extend the present single-UAV centralized design to multi-UAV cooperative ocean monitoring with distributed sensing and predictive beamforming. Such cooperation can provide spatial diversity and improve the overall robustness of predictive beamforming in practical maritime environments, especially when some buoys are partially blocked or intermittently obscured by sea waves.
	
	\vspace*{-2mm}
	\appendix
	
	\vspace*{-2mm}
	\subsection{Derivation of $\bar{\pmb{Q}}_{\text{s}}$}\label{App:DC}
	
	The discrete-time process noise $\bar{\pmb{u}}_{\text{s}}^{x(y)}$ represents the accumulated effect of $z_{\text{a}}^{x(y)}(t)$ over one sampling interval. For notational simplicity, the superscript `$x(y)$' is omitted. Following \cite{singer2007estimating}, the covariance matrix of $\bar{\pmb{u}}_{\text{s}}$ is given by
	\begin{align*} 
		\bar{\pmb{Q}}_{\text{s}} &= \mathbb{E}[\bar{\pmb{u}}_{\text{s}}\bar{\pmb{u}}_{\text{s}}^{\text{H}}]= 2\varsigma\sigma_a^2\int_0^{\bar{T}}  (e^{\pmb{\Phi}_{\text{s}}\tau} \pmb{e}_{\text{s}}) \cdot (\pmb{e}_{\text{s}}^{\text{T}} e^{\pmb{\Phi}^{\text{T}}_{\text{s}}\tau}) d \tau 
	\end{align*}
	\begin{align}
		&=\begin{bmatrix}
			q_{11} & q_{12} & q_{13} \\
			q_{12} & q_{22} & q_{23} \\
			q_{13} & q_{23} & q_{33}
		\end{bmatrix},
	\end{align}
	where the entries of $\bar{\pmb{Q}}_{\text{s}}$ are formulated as
	\begin{align} 
		q_{11} =& \frac{2 \bar{T}^3 \sigma_a^2}{3 \varsigma}-\frac{2 \bar{T}^2 \sigma_a^2}{\varsigma^2}+\frac{2 \bar{T} \sigma_a^2}{\varsigma^3}-\frac{4 \bar{T} \sigma_a^2}{\varsigma^3} e^{-\varsigma \bar{T}} \nonumber \\
		& +\frac{\sigma_a^2}{\varsigma^4}\left(1-e^{-2 \varsigma \bar{T}}\right), \\
		q_{12} =& \frac{\sigma_a^2}{\varsigma^3}\left(\bar{T}^2 \varsigma^2-2 \bar{T} \varsigma+1-2(1-\bar{T} \varsigma) e^{-\varsigma \bar{T}}+e^{-2 \varsigma \bar{T}}\right), \\ 
		q_{22} =& \frac{2 \bar{T} \sigma_a^2}{\varsigma}-\frac{\sigma_a^2}{\varsigma^2}\left(3-4 e^{-\varsigma \bar{T}}+e^{-2 \varsigma \bar{T}}\right), \\ 
		q_{23} =& \frac{\sigma_a^2}{\varsigma}\left(1-e^{-\varsigma \bar{T}}\right)^2,\nonumber \\  q_{13}&=\frac{\sigma_a^2}{\varsigma^2}\left(1-e^{-2 \varsigma \bar{T}}\right)-\frac{2 \bar{T} \sigma_a^2}{\varsigma} e^{-\varsigma \bar{T}}, \\ 
		q_{33} =& \sigma_a^2\left(1-e^{-2 \varsigma \bar{T}}\right).
	\end{align}
	
	\vspace*{-1mm}
	\subsection{Derivation of $\pmb A_{i,k,n}$, $i=1,2,3$}\label{App:rkn}
	
	\subsubsection{$\pmb A_{1,k,n}$}
	Based on the definition of $\pmb{H}_{k,n}^{(\pmb{r})}$, we have
	\begin{align} 
		\pmb{H}_{k,n}^{(\pmb{r})} = \bigg[\frac{\partial \pmb{r}_{k,n}}{\partial c_{k,n}^{x}},\pmb{0},\pmb{0},\frac{\partial \pmb{r}_{k,n}}{\partial c_{k,n}^{y}},\pmb{0},\pmb{0}\bigg].
	\end{align}
	To obtain $\frac{\partial \pmb{r}_{k,n}}{\partial c_{k,n}^{x(y)}}$, we apply the chain rule and first derive the partial derivatives of $\beta_{k,n}$, $\beta_{k,n}^{\text{clu}}$, and $\pmb{a}_{k,n}$ with respect to $c_{k,n}^{x}$ and $c_{k,n}^{y}$ as
	\begin{align} 
		& \frac{\partial \beta_{k,n}}{\partial c_{k,n}^{x(y)}} = \frac{\chi_{k,n} \Delta_{k,n}^{x(y)}}{2d_{k,n}^3}, \\
		& \frac{\partial \beta_{k,n}^{\text{clu}}}{\partial c_{k,n}^{x(y)}}\! =\! \frac{\Delta^{x(y)}_{k,n}}{\bar{d}_{k,n}}\! \left(\! \Xi_1 \frac{\Delta^z_{k,n}}{\bar{d}_{k,n}^2}\! +\! \Xi_2\Xi_{\text{e}} \bigg(\frac{\big(\Delta^z_{k,n}\big)^2}{d_{k,n} \bar{d}_{k,n}^2}\! +\! \frac{2 \Gamma_s d_{k,n}}{\big(\Delta^z_{k,n}\big)^2}\bigg)\! \right)\! , \\
		& \frac{\partial \pmb{a}_{k,n}}{\partial c_{k,n}^{x(y)}} = -j(\eta_{x(y)}\pmb{B}_{x(y)}+\eta_{xy}\pmb{B}_{y(x)})\pmb{a}_{k,n},
	\end{align}
	where $\Xi_{\text{e}}=e^{-\Gamma_\text{s}(\frac{\bar{d}_{k,n}}{c^z_{\text{U},n}})^2}$, $\pmb{B}_{x}= \pmb{\Lambda}_R \otimes \pmb{I}_R$, $\pmb{B}_{y}=\pmb{I}_R \otimes \pmb{\Lambda}_R $ with $\pmb{\Lambda}_R=\text{diag}(0,1,\dots,R-1)$, and 
	\begin{align} 
		\eta_{x(y)} =& \frac{\partial \mu_{k,n}^{x(y)}}{\partial c_{k,n}^{x(y)}}= \frac{2\pi  d_{\text{a}}}{\lambda }\frac{(\Delta_{k,n}^{y(x)})^2+(\Delta_{k,n}^{z})^2}{d_{k,n}^3}, \\
		\eta_{xy} =& \frac{\partial \mu_{k,n}^{x}}{\partial c_{k,n}^{y}}= \frac{-2\pi  d_{\text{a}}}{\lambda }\frac{\Delta_{k,n}^{x}\Delta_{k,n}^{y}}{d_{k,n}^3} .
	\end{align}
	
	Substituting the above relations yields 
	\begin{align} 
		\frac{\partial \pmb{r}_{k,n}}{\partial c_{k,n}^{x(y)}}=
		\frac{\partial \beta_{k,n}}{\partial c_{k,n}^{x(y)}} \pmb{a}_{k,n}+\beta_{k,n}	\frac{\partial \pmb{a}_{k,n}}{\partial c_{k,n}^{x(y)}}+\rho \frac{\partial \beta_{k,n}^{\text{clu}}}{\partial c_{k,n}^{x(y)}} \pmb{a}_{k,n}^{\text{clu}} ,
	\end{align}
	where $\pmb{a}_{k,n}^{\text{clu}}=\pmb{a}_{\text{rx}}(\hat{\pmb{c}}_{k,n|n-1},\pmb{c}_{\text{U},n})$. After computing $(\pmb{H}_{k,n}^{(\pmb{r})})^{\text{H}} \pmb{H}_{k,n}^{(\pmb{r})}$, $\pmb A_{1,k,n}$ is obtained.
	
	\subsubsection{$\pmb A_{2,k,n}$}
	From \eqref{eqfmea}, the Doppler-frequency measurement is
	\begin{align} 
		f_{k,n}^{\text{D}} = \frac{2}{\lambda}\left(\frac{\Delta_{k,n}^{x}}{d_{k,n}}v_{k,n}^x+\frac{\Delta_{k,n}^{y}}{d_{k,n}}v_{k,n}^y\right).
	\end{align}
	Then, the derivatives with respect to $\bar{v}_{k,n}^{x(y)}$ and $c_{k,n}^{x(y)}$ are derived as
	\begin{align} 
		\frac{\partial f_{k,n}^{\text{D}}}{\partial \bar{v}_{k,n}^{x(y)}} =& \frac{2}{\lambda}\frac{\Delta_{k,n}^{x(y)}}{d_{k,n}}, \\
		\frac{\partial f_{k,n}^{\text{D}}}{\partial c_{k,n}^{x(y)}} =& \frac{-2}{\lambda d_{k,n}^3 }\Big(v_{k,n}^{x(y)}((\Delta_{k,n}^{y(x)})^2+(\Delta_{k,n}^{z})^2) \nonumber \\
		& - v_{k,n}^{y(x)}\Delta_{k,n}^{x}\Delta_{k,n}^{y}\Big) .
	\end{align}
	According to the state ordering, the Jacobian row vector is
	\begin{align} 
		\pmb{h}_{k,n}^{({f })}=\bigg[\frac{\partial f_{k,n}^{\text{D}}}{\partial c_{k,n}^{x}},\frac{2}{\lambda}\frac{\Delta_{k,n}^{x}}{d_{k,n}},0,\frac{\partial f_{k,n}^{\text{D}}}{\partial c_{k,n}^{y}},\frac{2}{\lambda}\frac{\Delta_{k,n}^{y}}{d_{k,n}},0\bigg].
	\end{align}
	Evaluating $(\pmb{h}_{k,n}^{({f})})^{\text{H}} \pmb{h}_{k,n}^{({f})}$ then yields $\pmb A_{2,k,n}$.
	
	\subsubsection{$\pmb A_{3,k,n}$}
	As shown in \eqref{eqfmea}, the delay measurement is only associated with $c_{k,n}^{x(y)}$, and hence the corresponding Jacobian row vector is
	\begin{align} 
		\pmb{h}_{k,n}^{({\tau })} = 
		\bigg[\frac{-2}{c_0}\frac{\Delta_{k,n}^{x}}{d_{k,n}},0,0,\frac{-2}{c_0}\frac{\Delta_{k,n}^{y}}{d_{k,n}},0,0\bigg] .
	\end{align}
	After computing $(\pmb{h}_{k,n}^{({\tau})})^{\text{H}} \pmb{h}_{k,n}^{({\tau})}$, $\pmb A_{3,k,n}$ is obtained.
	
	\bibliographystyle{IEEEtran}
	\bibliography{UUSW}

\end{document}